\documentclass{emulateapj}

\shorttitle{Vertical Distributions of Resolved Stars}
\shortauthors{Seth, Dalcanton \& de Jong}

\begin{document}

\slugcomment{Accepted by the Astronomical Journal, June 4, 2005.}

\title{A Study of Edge-On Galaxies with the Hubble Space Telescope's
  Advanced Camera for Surveys. II. Vertical distribution of the resolved stellar population}

\author{Anil C. Seth}
\affil{University of Washington}
\email{seth@astro.washington.edu}

\author{Julianne J. Dalcanton\footnote{Alfred P. Sloan Research Fellow}}
\affil{University of Washington}
\email{jd@astro.washington.edu}

\author{Roelof S. de Jong}
\affil{Space Telescope Science Institute}
\email{dejong@stsci.edu}

\begin{abstract}

We analyze the vertical distribution of the resolved stellar
populations in six low-mass (V$_{\rm max} = 67 - 131$~km~s$^{-1}$),
edge-on, spiral galaxies observed with the Hubble Space Telescope
Advanced Camera for Surveys.  In each galaxy we find evidence for an
extraplanar stellar component extending up to 15 scale heights (3.5
kpc) above the plane, with a scale height typically twice that of 2-D
fits to $K_s$ band 2MASS images.  We analyze the vertical distribution
as a function of stellar age by tracking changes in the
color-magnitude diagram.  The young stellar component
($\lesssim$10$^8$ yrs) is found to have a scale height larger than the
young component in the Milky Way, suggesting that stars in these low
mass galaxies form in a thicker disk.  We also find that the scale
height of a stellar population increases with age, with young main
sequence stars, intermediate age asymptotic giant branch stars, and
old red giant branch stars having succesively larger scale heights in
each galaxy.  This systematic trend indicates that disk heating must
play some role in producing the extraplanar stars.  We constrain the
rate of disk heating using the observed trend between scale height and
stellar age, and find that the observed heating rates are dramatically
smaller than in the Milky Way.  The color distributions of the red
giant branch stars well above the midplane indicate that the extended
stellar components we see are moderately metal-poor, with peak
metallicities around [Fe/H]$=-1$ and with little or no metallicity
gradient with height.  The lack of metallicity gradient can be
explained if a majority of extraplanar RGB stars were formed at early
times and are not dominated by a younger heated population.  Our
observations suggest that, like the Milky Way, low-mass disk galaxies
also have multiple stellar components.  In its structure, mean
metallicity and old age, the RGB component in these galaxies seems
analagous to the Milky Way thick disk. However, without additional
kinematic \& abundance measurements, this association is only
circumstantial, particularly in light of the clear existence of some
disk heating at intermediate ages.  Finally, we find that the vertical
dust distribution has a scale height somewhat larger than that of the
main sequence stars.

\end{abstract}
\keywords{galaxies:spiral -- galaxies:structure -- galaxies:formation -- galaxies:individual(IC~5052, NGC~55, NGC~4144, NGC~4244, NGC~4631, NGC~5023) -- dust}

\section{Introduction}

Galactic structure has long been recognized as a key constraint on
theories of galaxy formation. The presence of two distinct components
-- disks and spheroids -- suggests that at least two separate physical
mechanisms were active during the formation of spiral galaxies.
Within the Milky Way, however, detailed analyses of the kinematics and
chemistry of nearby stars have revealed additional components: a thin
disk with scale heights dependent on age, a thick disk and a
stellar halo.  Each of these components place unique constraints on
the Galaxy's evolution.  In particular, the ages and metallicities of
the old thin disk, thick disk and halo indicate that they are remnants
of the initial stages in the assembly of the galaxy and thus
constrain the early evolution of galaxies.

Several scenarios for the creation of the old thin and thick disks are
consistent with current observations in the Milky Way.  These
scenarios can be categorized into three main types: (1) the creation
of a thick disk from a thin disk by heating by molecular clouds,
spiral arms, star formation or accretion events
\citep[e.g.][]{spitzer51,barbanis67,lacey91,kroupa02,quinn93,gnedin03}.
(2) the slow collapse of the proto-Galaxy forming the thick and thin
disk in succession \citep[e.g.][]{eggen62,gilmore84}, and (3) the
formation of a thick disk from mergers, either by direct accretion of
stars or by {\it in situ} formation from accreted gas
\citep[e.g.][]{bekki01,gilmore02,abadi03,brook04}.  In this last
scenario the thin disk would likely form by a settling of gas from the
merger events.  It is widely assumed that the stellar halo forms from
accreted satellites \citep[see review by][]{freeman02}.  Determination
of which processes are important in galaxy formation requires the
study of stellar components in other galaxies.

Unfortunately, detailed analyses of older stellar components are
difficult outside the Milky Way.  Some thick disks and stellar halos
have been identified using broad-band surface photometry
\citep[e.g.][]{burstein79,tsikoudi79,dalcanton02,pohlen04,zibetti04a},
but their low surface brightness precludes all but a most cursory
study of their structure and stellar content.  Recently, HST imaging
has begun to allow richer analyses of the resolved stars in the thick
disk and halo \citep{brown03,tikhonov05a,mould05}.  The resulting
color-magnitude diagrams can provide robust constraints on the
metallicities and ages of the stars which dominate these old
components in other galaxies.

In this paper we take advantage of a large database of ACS imaging of
nearby edge-on galaxies to analyze the stellar populations of the
galaxies as a function of height above the galaxy midplane (``disk
height'').  The wide field of view of ACS allows us to track changes
in the color-magnitude diagram to more than 3 kpc above the plane.
Moreover, our sample galaxies have much lower masses than the Milky
Way, and thus constrain how the extraplanar stellar content varies
with galaxy mass.  The properties of the sample and their global
color-magnitude diagrams are given in \citet{seth05} (Paper I).  In
this paper, we study the spatially resolved stellar populations in
eight fields of six galaxies that were close enough to determine
accurate Tip of the Red Giant Branch (TRGB) distances.  In particular,
we show that there is a significant stellar population high above the
plane of these galaxies and that these stars are old and metal-poor,
comparable to high-latitude stellar populations in the Milky Way.

In \S2, we review the results of Paper I for these six galaxies
and describe the completeness corrections used to correct the stellar number
counts in the remainder of the paper.  In \S3 we show
that we trace stars in the host galaxies out to many scale heights and
that these stars have a larger scale height than expected from fits to
2MASS $K_s$ band images.  In \S4 we select three populations of stars
from color-magnitude diagrams (CMDs) and show that the scale height of
a stellar population increases with age.  We
derive metallicity distribution functions for the extended stellar
component in \S5 and then conclude by discussing our results in a
broader context in \S6.

\section{Review of the Galaxy Properties}

\begin{deluxetable*}{lcccccccccc}
%\rotate
\tablewidth{0pt}
\tablecaption{Galaxy Sample Properties}
\tablehead{
     \colhead{Galaxy}  &
     \colhead{RA(J2000)} &
     \colhead{Dec(J2000)} &
     \colhead{$incl.$} &
     \colhead{Type} &
     \colhead{$V_{max}$} &
     \colhead{$m-M_{\rm TRGB}$} &
     \colhead{$h_{R}$} &
     \colhead{$h_{R}$} &
     \colhead{$z_{1/2}$} &
     \colhead{$z_{1/2}$} \\
     \colhead{}  &
     \colhead{} &
     \colhead{} &
     \colhead{} &
     \colhead{} &
     \colhead{km/sec} &
     \colhead{} &
     \colhead{[\arcsec]} &
     \colhead{[kpc]} &
     \colhead{[\arcsec]} &
     \colhead{[pc]} 
} 

\startdata
IC 5052         & 20 52 02.9  & -69 11 45 & 89   & SBcd & 79  & 28.90 & 53.87 & 1.57 &  7.33 & 214 \\ 
NGC 55          & 00 14 54.4  & -39 11 59 & 80   & SBm  & 67  & 26.63 & 93.30 & 0.96 & 23.34 & 240 \\ 
NGC 4144        & 12 09 58.3  & +46 27 28 & 83   & SBc  & 67  & 29.36 & 30.56 & 1.10 &  6.96 & 251 \\ 
NGC 4244        & 12 17 29.5  & +37 48 28 & 84.5 & Sc   & 93  & 28.20 & 84.27 & 1.78 & 12.15 & 257 \\ 
NGC 4631        & 12 42 07.7  & +32 32 33 & 85   & SBcd & 131 & 29.42 & 35.49 & 1.32 &  7.54 & 280 \\ 
NGC 5023        & 13 12 11.7  & +44 02 17 & 88   & Sc   & 77  & 29.10 & 39.81 & 1.28 &  5.10 & 160 \\ 
\enddata
\tablecomments{RA, Dec, TRGB distance moduli, $h_r$, and $z_{1/2}$
  taken from Paper I.  Galaxy Types and $V_{\rm max}$ from HYPERLEDA/LEDA
  \citep{paturel95,paturel03}.  Inclinations taken from
  \citet{becker88,hummel86,martin98,olling96,hummel90,degrijs96} (in
  same order as table).   
}
\end{deluxetable*}

Table~1 shows the position, type, maximum circular velocity 
($V_{\rm max}$), distance modulus ($m-M_{\rm TRGB}$), scale length ($h_r$) and
the $K_s$ band half-light height ($z_{1/2}$) of the six galaxies we will
discuss in this paper.  The latter two parameters
were determined from 2-D model fits to 2MASS $K_s$ band data, presented
in Paper I.  The vertical component of these models is defined using
the distribution of an isothermal population of stars \citep{vanderkruit81}:
\begin{equation}
\Sigma(z) \propto {\rm sech}^{2}\left( \frac{z}{z_0} \right)
\end{equation}
where $\Sigma(z)$ is the surface brightness or density at a position $z$
above the midplane, and $z_0$ is the scale height.  We will use this
functional form to fit vertical distributions throughout this paper.
Note, however, that this is one of many equations commonly used to
describe the vertical distribution of stars in galaxies, a good
overview of which can be found in \citep{pohlen00}.  Most of these
functional forms vary near the midplane, but have similar exponential
declines at large disk heights.  When comparing galaxies in this
paper, the disk heights will be normalized by the $z_{1/2}$ parameter,
which gives the height containing 50\% of the $K_s$ band light in the
model fits.  It is related to $z_0$, $z_{1/2}$=0.549~$z_0$, and is
similar to the exponential scale height $h_z$, which at large scale
heights is equal to $\frac{1}{2} z_0$ \citep{vanderkruit81}.
The values of $z_{1/2}$ for the six galaxies
range from 160-280 pc.  For comparison, the Milky Way thin disk has
exponential scale heights ranging from $\sim$100 pc at young ages
\citep{schmidt63} to 330 pc \citep{chen01} at older ages.  

All six galaxies are within 8 Mpc and are
type Sc or later.  The maximum circular velocities are all below 135
km/sec suggesting that these objects are closer in mass to the LMC
than to the Milky Way.  We note that all the galaxies except NGC~4631
have circular velocities well below 120 km/sec, which appears to mark
a break in the dust properties \citep{dalcanton04} and current
metallicity \citep{garnett02} in spiral galaxies.
The scale lengths of our sample galaxies range from 0.9 to 1.6 kpc,
$\sim$2-3 times smaller than the Milky Way scale length\footnote{The Milky Way thin \& thick disk scale lengths are rather uncertain with recent values ranging between 2-4 kpc \citep[e.g.][]{ng96,mendez98,ojha01}, we will use a scale length of 3 kpc for comparisons to the Milky Way made in this work.}.  None of the
galaxies has an apparent bulge component, although NGC~4244 does have
a prominent central stellar cluster, which is clearly visible in the
ACS and 2MASS $K_s$ band images.   

Our observations include eight HST/ACS fields in the six galaxies
shown in Table~1 and are described fully in Paper~I \citep{seth05}.  These
observations were obtained to study the dust-lane properties of these
galaxies, and hence are centered on the galaxy midplane.  The
dimensions of these eight fields are given in Table~2, which shows the
minimum and maximum disk radius in terms of the scale length and the
minimum and maximum disk height in terms of $z_{1/2}$.  In general,
each field is located close to the center of the galaxy.  However, two
of the galaxies, NGC~55 and NGC~4631 have additional fields located
further out in the disk that are given a '-DISK' suffix.  Note that
many of the galaxies lie diagonally across the chip meaning that the
extremities of the ranges given in Table~2 are not well sampled.  In
this paper, we focus on the vertical distributions of the stars, and
where not otherwise noted, analyze all data at the same disk height
together.  

The approach proposed above is valid as long as the scale height of
disk components does not vary substantially with radius, an assumption
that has been verified through observations of edge-on galaxies
\citep[e.g.][]{vanderkruit81,pohlen04}.  We note however, that there
are some analyses which indicate that the scale height of galaxies
flares with increasing radius, both in our own Galaxy \citep{lopez02}
and in edge-on galaxies \citep{degrijs97,narayan02}.  However,
\citet{degrijs97} show that in their sample of edge-on galaxies,
late-type galaxies such as those observed here have little or no
flaring.  In one of our galaxies, NGC~4244, flaring of the HI gas by a
factor of $\sim$3 is seen between radii of 8-13 kpc \citep{olling96}.
Even if similar flaring occurs in the stellar distribution, the star
counts at the disk heights we consider here will still be dominated by
stars located (radially) near the center of the galaxy.  We note
that we do see some evidence for modest flaring in our two '-DISK' fields
which lie at large radii (see \S4.2.1). 

\begin{deluxetable}{lccccc}
\tablewidth{0pt}
\tablecaption{Field properties}
\tablehead{
     \colhead{Field}  &
     \colhead{\# Stars}  &
     \colhead{R$_{\rm min}$} &
     \colhead{R$_{\rm max}$} &
     \colhead{$z_{\rm min}$} &
     \colhead{$z_{\rm max}$} \\
     \colhead{}  &
     \colhead{}  &
     \colhead{[R/$h_r$]} &
     \colhead{[R/$h_r$]} &
     \colhead{[$z/z_{1/2}$]} &
     \colhead{[$z/z_{1/2}$]} 
}
\startdata
IC 5052        &  68636 &  -2.9 &  1.9 &  -22.1 & 15.6  \\ 
NGC 55         & 281536 &  -1.6 &  1.4 &   -5.8 &  5.3  \\ 
NGC 55-DISK    & 253108 &   3.3 &  6.2 &   -7.3 &  5.5  \\ 
NGC 4144       &  60552 &  -2.3 &  5.9 &  -18.7 & 14.6  \\ 
NGC 4244       & 121238 &  -1.4 &  1.2 &   -6.8 & 10.5  \\ 
NGC 4631       & 104940 &  -3.1 &  3.4 &  -17.8 & 10.2  \\ 
NGC 4631-DISK  &  97656 & -10.1 & -1.6 &  -21.2 & 14.8  \\ 
NGC 5023       &  42293 &  -2.7 &  3.1 &  -18.1 & 25.3  \\ 
\enddata
\end{deluxetable}

In Paper I we presented the color-magnitude diagrams (CMDs) for each of
the galaxy fields discussed here.  These are reproduced in Figure~1
which shows the F606W-F814W color vs. the F814W magnitude.  All
magnitudes in this paper are given in the VEGAmag system (see Paper I
for details).  Each of the CMDs show plumes at F606W-F814W colors
of $\sim$0.1 and $\sim$1.3 which are young main sequence (MS) and
helium burning (HeB) stars respectively.  An old RGB can
be seen in each galaxy at colors of $\sim$1 and an intermediate-age
asymptotic giant branch (AGB) extends from the tip of the RGB to
redder colors.  Further details on the components can be found in
\S4.1 and in Paper I.  The boxes shown in Figure~1 isolating these
components will be discussed in greater detail in \S4.

\begin{figure*}
\plotone{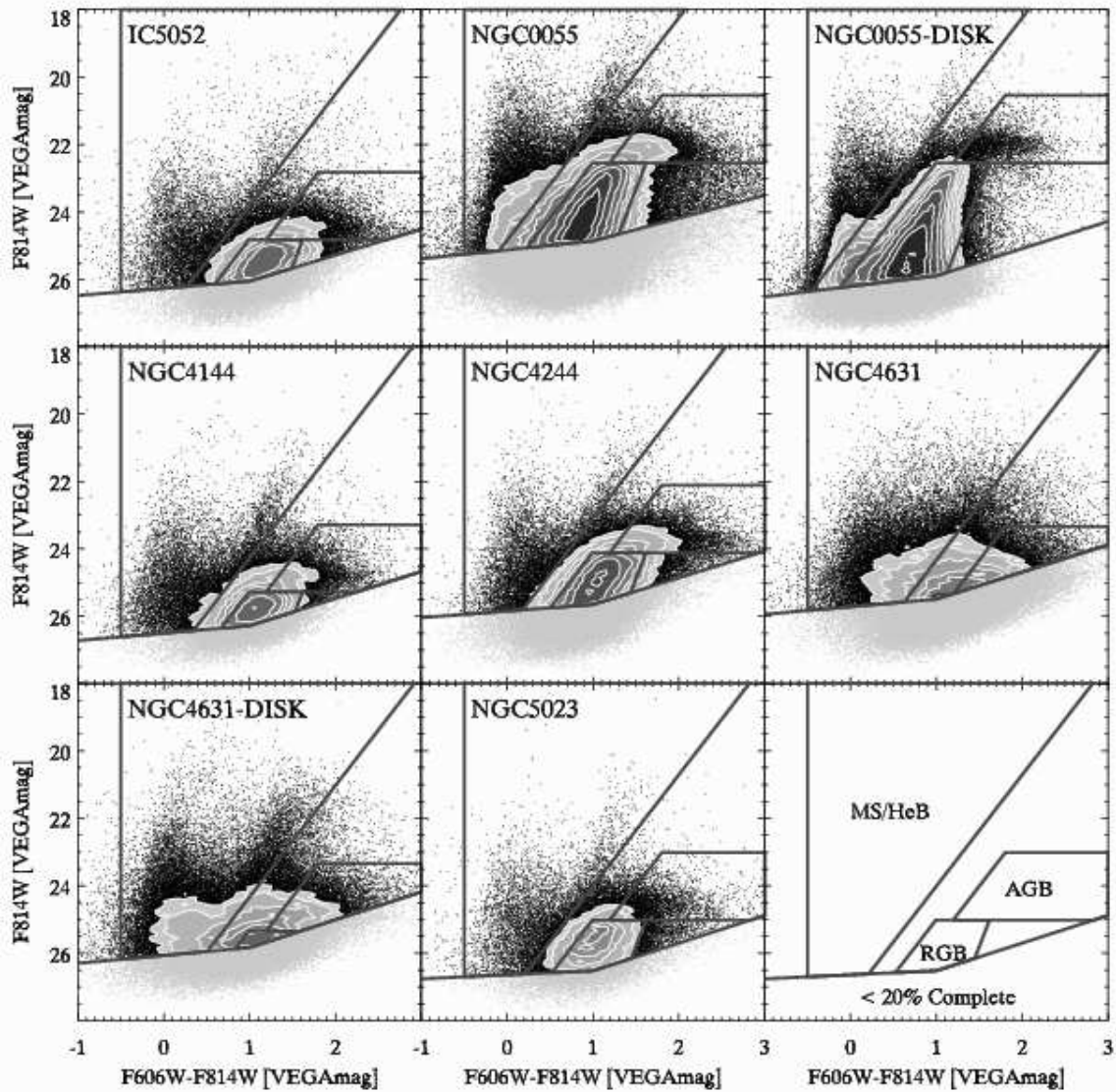}
\caption{Color-magnitude diagrams of each galaxy. Contours are used
  where the density of points becomes high.  The contours are drawn at  
  densities of 75, 100, 150, 200, 250, 350, 500, 750, 1000, and 1500
  stars per 0.1 magnitude and color bin.  The bottom-left panel
  contains a key to the gray lines which demarcate sections of the
  CMD.  The line across the bottom shows the 20\% completeness limit
  in bright regions of the galaxy (see \S2.1 for details) and the
  other lines show the regions used in each galaxy for the MS/HeB, AGB
  and RGB stars.} 
\end{figure*}

\subsection{Completeness Corrections}

As described in Paper I, we conducted extensive artificial star tests
to characterize the completeness in each field as
a function of magnitude, color and location.  The galaxies in our
sample have high surface brightnesses near their midplane making the
completeness there much lower (see Fig.~2 from Paper~I).  The goal of
this paper is to analyze the vertical distribution of stellar
populations, and therefore correcting for this varying completeness is
critical. 

For each field, over 5 million artificial stars were inserted at
random positions, with magnitudes between F606W of 18 and 29, and with
F606W-F814W colors between -1 and 3.  These stars were then run
through the same pipeline used to determine the stellar photometry.
Artificial stars that coincided with actual sources were considered
detected only if the input magnitude of the artificial star was
brighter than the actual source in both bands.

To enable completeness corrections for individual stars, we determined
our completeness for the artificial stars in bins of magnitude, color
and local surface brightness.  For the magnitude and color, we used
0.15 magnitude wide bins.  At its steepest, the completeness as a
function of magnitude varies by at most 6\% over 0.15 magnitudes, so
any error introduced by the binning should be smaller than that.  We
then determined the size of the surface brightness bins such that
there would be $\sim$50 stars in each final bin.  Determining the
completeness from 50 stars gives a random error in the completeness of
$\sim$6\%.  An aperture around each star from 11 to 14 pixels was used
to determine the local surface brightness.  We determined the surface
brightness level from the F606W image so as to include the effects of
the HII region emission visible in a number of galaxies (most notably
NGC 55, which therefore has the brightest completeness level in
Fig.~2).  This emission is not seen at all in the F814W images, due to
the lack of strong emission lines in F814W bandpass.  Using this
binned completeness function, we are able to determine the
completeness of any individual star based on its color, magnitude and
local surface brightness level to within $\sim$10\%.  In the stellar
density profiles presented in \S3 \& \S4, the completeness corrections
are up to 200\% near the midplane, but fall to $\lesssim30\%$ at
$z/z_{1/2} > 3$.

In addition to correcting for the completeness, magnitude limits must
also be set to insure that we are not using stars fainter than we can
detect in the higher surface brightness regions of the image (i.e. the
midplane).  We therefore choose to limit our analysis to regions of
the CMD that fall above a conservative 20\% completeness limit in
regions of high surface brightness.  Figure~2 shows the 20\%
completeness limits for the brightest regions in each field.  As can
be seen, the completeness limit rises towards redder colors, and
steepens at colors redder than F606W-F814W of 1.  To make a smooth
boundary that is easily applied to our data, we fit the 20\%
completeness curves to two lines intersecting at F606W-F814W=1.
Table~2 shows the results of these fits, by giving the completeness
limit at F606W-F814W of -1 (F814W$_{\rm lim,-1}$), 1 (F814W$_{\rm
lim,1}$), and 3 (F814W$_{\rm lim,3}$).  We use these limits throughout
this paper to insure that comparisons made between stellar populations
at different disk heights are valid.

We note here that although we can correct for incompleteness due to
crowding, we cannot correct for the attenuation of stars by dust.  We
will show in \S4 that all the galaxies in our sample are optically
thick near their midplanes.  Therefore at low galactic latitudes, our
completeness corrected stellar census will fall short of the true
number of stars.

\begin{figure}
\plotone{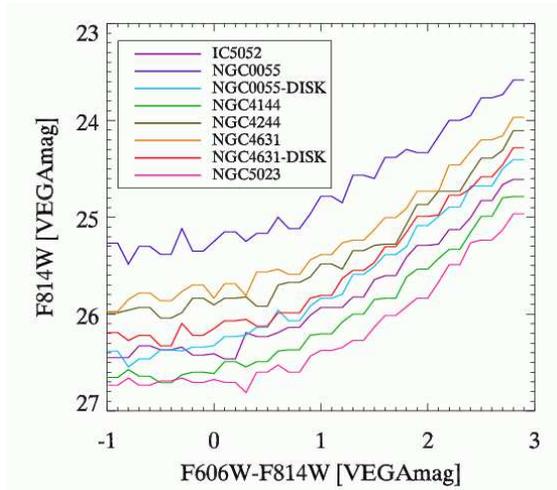}
\caption{The 20\% completeness limit in the brightest sections of each
  field as a function of color.  The variation from field-to-field
  is due to increased crowding / higher surface brightness regions in
  some fields.}
\end{figure}

\begin{deluxetable}{lccc}
\tablewidth{0pt}
\tablecaption{20\% Completeness Limits}
\tablehead{
     \colhead{Field}  &
     \colhead{F814W$_{\rm lim,-1}$} &
     \colhead{F814W$_{\rm lim,1}$} &
     \colhead{F814W$_{\rm lim,3}$} \\
     \colhead{}  &
     \colhead{[mag.]} &
     \colhead{[mag.]} &
     \colhead{[mag.]} 
}
\startdata
IC 5052         & 26.48 & 26.05 & 24.51  \\ 
NGC 55          & 25.39 & 24.86 & 23.51  \\ 
NGC 55-DISK     & 26.52 & 25.89 & 24.28  \\ 
NGC 4144        & 26.72 & 26.28 & 24.67  \\ 
NGC 4244        & 26.04 & 25.65 & 24.07  \\ 
NGC 4631        & 25.94 & 25.49 & 23.88  \\ 
NGC 4631-DISK   & 26.30 & 25.83 & 24.18  \\ 
NGC 5023        & 26.76 & 26.52 & 24.88  \\ 
\enddata
\end{deluxetable}

\section{Vertical Distribution of Stars}

We demonstrate in this section that there are significant numbers of
stars well above the planes of all our disks and that the profiles of
these stars do not fit the profiles expected from the ground-based
$K_s$ band galaxy fits.

\begin{figure}
\plotone{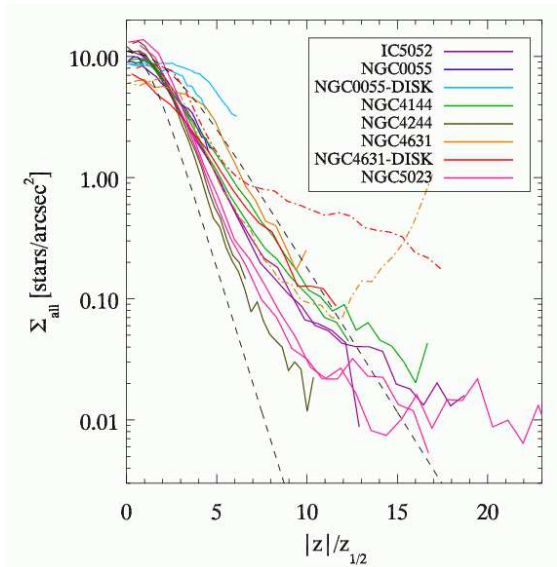}
\caption{The surface density of stars as a function of height from the
  midplane.  Curves have been completeness corrected as described in
  the text.  Each galaxy has two lines, one for above and below the
  plane.  The dashed lines indicate stellar distributions with scale
  heights 1 \& 2 $\times$ that of the $K_s$ band fits to the disks.
  The dot-dashed lines for the NGC~4631 fields indicate the regions
  where contamination of stars from companion NGC~4627 is
  significant.}
\end{figure}

\begin{deluxetable*}{lccccccccc}
%\rotate
\tablewidth{0pt}
\tablecaption{Resolved Stellar Component Scale Heights}
\tablehead{
     \colhead{Field}  &
     \colhead{$K_s$} & 
     \colhead{All $+$} &
     \colhead{All $-$} &
     \colhead{MS} &
     \colhead{AGB} &
     \colhead{RGB} &
     \colhead{$h_{r}/z_{\rm 0,MS}$} &
     \colhead{$h_{r}/z_{\rm 0,AGB}$} &
     \colhead{$h_{r}/z_{\rm 0,RGB}$} \\
     \colhead{}  &
     \colhead{$z_0$ [pc]}  &
     \colhead{$z_0$ [pc]}  &
     \colhead{$z_0$ [pc]}  &
     \colhead{$z_0$ [pc]}  &
     \colhead{$z_0$ [pc]}  &
     \colhead{$z_0$ [pc]}  &
     \colhead{}  &
     \colhead{}  &
     \colhead{}  
}
\startdata
IC 5052\tablenotemark{a}      & 390 & 767$\pm$30 & 686$\pm$44 &  261$\pm$11 & 467$\pm$6  & 655$\pm$6    & 6.0 & 3.4 & 2.4 \\ 
NGC 55                        & 437 &            &            &  327$\pm$24 & 644$\pm$10 & 701$\pm$3    & 2.9 & 1.5 & 1.4 \\ 
NGC 55-DISK                   & 437 &            &            &  526$\pm$1  & 741$\pm$15 & 999$\pm$6    & 1.8 & 1.3 & 1.0 \\ 
NGC 4144                      & 457 & 940$\pm$41 & 965$\pm$15 &  374$\pm$27 & 699$\pm$16 & 934$\pm$18   & 2.9 & 1.6 & 1.2 \\ 
NGC 4244                      & 468 & 740$\pm$40 &            &  325$\pm$20 & 443$\pm$24 & 551$\pm$9    & 5.5 & 4.0 & 3.2 \\ 
NGC 4631\tablenotemark{b}     & 510 & 927$\pm$46 &            &  510$\pm$26 & 895$\pm$51 & 1154$\pm$194 & 2.6 & 1.5 & 1.1 \\ 
NGC 4631-DISK\tablenotemark{b}& 510 &1131$\pm$50 &            &  505$\pm$22 & 1200$\pm$1 & 1387$\pm$73  & 2.6 & 1.5 & 1.0 \\ 
NGC 5023\tablenotemark{a}     & 291 & 505$\pm$32 & 534$\pm$26 &  204$\pm$6  & 289$\pm$6  & 391$\pm$4    & 6.3 & 4.4 & 3.3 \\ 
\enddata
\tablecomments{{\it (a)} In IC~5052 and NGC~5023, MS, AGB and RGB fits
  excluded disk heights below 1.5$z_{1/2}$.  All other galaxies
  excluded below 3$z_{1/2}$. {\it (b)} Negative values of $z$ excluded
  beyond 2 kpc due to the presence of companion galaxy NGC 4627.}  
\end{deluxetable*}

In Figure~3 we present the completeness corrected surface density
profiles of all the detected stars ($\Sigma_{\rm all}$) above the
completeness limits given in Table~3.  Two lines are shown for each
galaxy, giving the profile on both sides of the disk.  To determine
the surface density, stars were binned as a function of scale height.
After binning, the completeness-corrected number of stars was divided
by the area of the bin to obtain the surface density.  Only bins with an
area of more than 300 arcsec$^2$ are plotted.

Figure~3 shows that we trace the stellar component of each galaxy out
to large disk heights, with several galaxies being traced out beyond
10$z_{1/2}$.  The profiles in general are fairly symmetric, the most
notable exception being NGC~4631 and NGC~4631-DISK.  These fields are
contaminated on one side (shown using dot-dashed lines) by the presence of
the companion galaxy NGC 4627 (see Fig.~3 from Paper~I).  The decrease
of the profiles with increasing scale height out to the edge of the
fields strongly suggests that the profiles remain above the surface
density of foreground Galactic stars and background unresolved dwarf
galaxies in our magnitude range.  Only above 10$z_{1/2}$ in IC~5052
and NGC~5023 is there some evidence for the leveling off that would be
expected as we reach the foreground/background level.  Note that the
galaxies are all at galactic latitudes above 35$^\circ$.  We therefore
can safely assume that a vast majority of stars in our images are
located in the host galaxies and we make no corrections for
foreground/background sources.

The dashed lines in Figure~3 show sech$^2$ profiles with scale heights
one and two times the measured $K_s$ band scale height (note that
because the plot is scaled by the $K_s$ band scale height these
profiles are the same for each galaxy).  As described in detail in
Paper I, models were fit using a Levenberg-Marquardt algorithm with
uniform weighting on all unmasked pixels to $K_s$ band 2MASS data.
Because the 2MASS surface photometry has a limiting isophote of $K_s
\sim 20.0$ mag/arcsec$^2$, only relatively high surface brightness
features could be fit.  The typical $K_s$ band peak surface brightness
of our galaxies is $\sim$18 (Paper~I, Table~2), which means that
galaxies are only detected in the $K_s$ images out to a few $z_{1/2}$
from the midplane.  Although the $K_s$ band light is often thought to
be dominated by the RGB stars that trace an old stellar population, we
will show in \S4 that in these low mass, late-type galaxies, it more
closely traces the young and intermediate age populations and is thus
dominated by red supergiant and AGB stars.

Figure~3 shows that the outer portions of the stellar density profiles
of the galaxies appear to be broader than indicated by the $K_s$ band
scale height.  To quantify this, we fit the stellar density profiles
between 5$z_{1/2}$ and 10$z_{1/2}$ on each side of the midplane in
each galaxy.  Only profiles with data beyond 8$z_{1/2}$ were fit (thus
excluding NGC~55, NGC~55-DISK and one side of NGC~4244).  In the Milky
Way, the stellar profile deviates from the thin disk profile
beyond $\sim$1 kpc above the plane \citep{gilmore83}.  For our galaxies,
5$z_{1/2}$ is $\sim$1 kpc.  Therefore we would expect our fitting
range to be sensitive to a possible thick disk component in these
galaxies.  The scale heights of these fits above and below the
midplane are shown in the third and fourth columns (respectively) of
Table~4.  All of the fitted stellar profiles are significantly broader
($\times$1.5-2.4) than the $K_s$ band scale height.  This observation
strongly suggests that these galaxies contain a more
broadly distributed stellar component not traced by the $K_s$ band
2MASS images.

We note that the fitted scale heights, both in the $K_s$ band
and using stellar density profiles, can differ from the true scale
height due to a number of factors.  First, dust attenuation can
obscure light near the midplane of the galaxy. In the $K_s$ band this
attenuation should be small, and we ameliorate this problem in the fits
presented here by avoiding the midplane.  Second, the galaxies may not be
perfectly edge-on. Based on previous observations, the least inclined
of the galaxies is NGC~55, which has a $\sim$80$^\circ$ inclination
\citep{hummel86}.  This would give a fitted scale height $\sim$30\%
greater than the intrinsic disk scale height.  
We also note that NGC~55 and NGC~4631 are fairly irregularly shaped
making the fits to these galaxies less reliable than for the other
four galaxies. 

\section{Variation in distribution with stellar population}

We now turn our analysis to stars selected in regions of our CMDs that
isolate stellar populations with different ages.  Using this method we
show that the older stellar populations have an increasing scale
height.  In \S4.3, we examine the variation of scale height in the
context of disk heating models.  We then present simplistic dust
models in \S4.4 and show that the dust extinction in these galaxies is
distributed in a component that is broader than the young stellar
populations.

\subsection{Selection of CMD regions}

We attempt to separate our data into young, intermediate, and old
stellar populations by selecting stars from different regions in the
CMD.  The young stars are found in the Main Sequence (MS) component
and in the red and blue Helium Burning (RHeB,BHeB) sequences (see
Paper I, Fig.~1 for a schematic CMD), all of which should contain
stars under a few 100 Myr in age. For the intermediate age stars we
select AGB stars brighter than and redward of the RGB, resulting in
ages ranging between a few 100 Myr and a few Gyr.  Lastly, for the old
population of stars we select RGB stars, which have ages in excess of
1 Gyr, although some AGB stars will also be found in the same region.

The actual regions used for the selection are shown in the CMDs in
Fig.~1.  The bottom right figure is a cartoon illustrating the
selection of the MS/HeB, AGB and RGB regions.  The RGB region was
selected using lines with slopes of 3.3 and 6.6 and F606W-F814W colors
of 1.0 and 1.6 at the tip of the RGB.  The MS region was defined by
taking all stars redward of -0.5 and blueward of a line with slope of
3.3 and a color of 0.7 at the TRGB magnitude.  Finally, the AGB region
isolates stars less than two magnitudes brighter than the TRGB
magnitude and redwards of a line with slope of 3.3 and a color of 1.2
at the TRGB magnitude.  These boundaries were combined with the TRGB
magnitude given in Table~1 and the completeness limits from Table~3 to
determine the final regions for each galaxy shown in Figure~1.  The
regions were chosen somewhat conservatively - e.g. we chose to put
space between the MS/HeB section and the RGB so that there would be
little overlap between the two due to dust extinction or large
photometric errors at faint magnitudes.  From here we will refer to
the stars in these CMD boxes as the MS, AGB and RGB stellar
populations.

\subsubsection{Synthetic CMDs}

To determine the typical ages of stars detected in our CMD boxes and
to facilitate quantitative comparisons between galaxies and their
different stellar populations, we generated synthetic CMDs using the
MATCH program \citep{dolphin02} and the IAC-STAR program\footnote{This
work has made use of the IAC-STAR Synthetic CMD computation code,
IAC-STAR is supported and maintained by the computer divison of the
IAC.}  \citep{aparicio04}, using isochrones of \citet{bertelli94} and
\citet{girardi00} in both cases. The synthetic stars were generated
assuming a constant star formation rate (SFR) from 13 Gyr ago to the
present, and a metallicity that steadily increased from [Fe/H]$ =
-1.7$ to -0.4 \citep{garnett02}.  We used slightly different IMFs in
the two CMDs.  For the MATCH CMD a pure Salpeter IMF ($\alpha = 2.35$)
is assumed between 0.1 and 120 M$_\odot$, whereas in the IAC-STAR CMD
we used the default \citet{kroupa93} IMF, which is steeper at the high
mass end ($\alpha = 2.70$).

To compare these CMDs to our observations, the synthetic stars were
first transformed from Johnson V \& I to VEGAmag F606W \& F814W
colors.  We then mimicked observations of each galaxy as follows.
First, each star was randomly assigned a surface brightness value
based on the values of detected stars in each frame.  Then, using the
artificial star tests, we determined the chance each star was detected
and a magnitude error based on the star's initial F606W \& F814W
magnitudes (assuming the distance moduli shown in Table~1) and the
surface brightness value.  A final CMD was then made by randomly
determining if each star was detected and applying the determined
errors.  The resulting CMDs looked qualitatively similar to our
observed CMDs, with the most notable difference being an offset of the
AGB stars to somewhat brighter magnitudes in both synthetic CMDs
relative to the real data and a deficit of MS/HeB stars in the
IAC-STAR CMDs relative to the MATCH and the real galaxy CMDs.  This
could indicate either that the galaxies have enhanced recent star
formation or that their IMF is not as steep as the \citet{kroupa93}
IMF on the high mass end.

Figure~4 shows the resulting age distribution of the MS, AGB, and RGB
boxes in NGC~4144 using the MATCH and IAC-STAR synthetic CMDs.  The
age distributions for other galaxies are similar.  This figure clearly
demonstrates that we are separating the stars into young,
intermediate-age and old populations with our chosen CMD boxes.
However, the separation is not perfect.  Each bin has significant
overlaps with the others due to unavoidable photometric errors and to
true overlap in the colors and magnitudes of stellar populations with
different ages. The IAC-STAR CMDs have age distributions similar to
the MATCH CMDs, but with the AGB populations weighted more towards
older ages and a more significant contamination of old stars in the MS
box (probably due to the relative lack of MS stars in the IAC-STAR
models).  Both effects likely result from the steeper IMF assumed for
the Aparicio CMD.  In the following sections of the paper, we use the
MATCH CMD for comparisons with observations because it more closely
reproduces the ratio of young MS and HeB stars relative to the number
of older stars.

We note that these sunthetic CMDs assume a {\it constant} star
formation rate and thus are not useful in determining true star
formation histories.  However, we will be able to use them to get a
sense of relative star formation histories (SFHs) as a function of
scale height and to get a rough sense of the ages of the stars in our
CMD regions.

\begin{figure*}
\plottwo{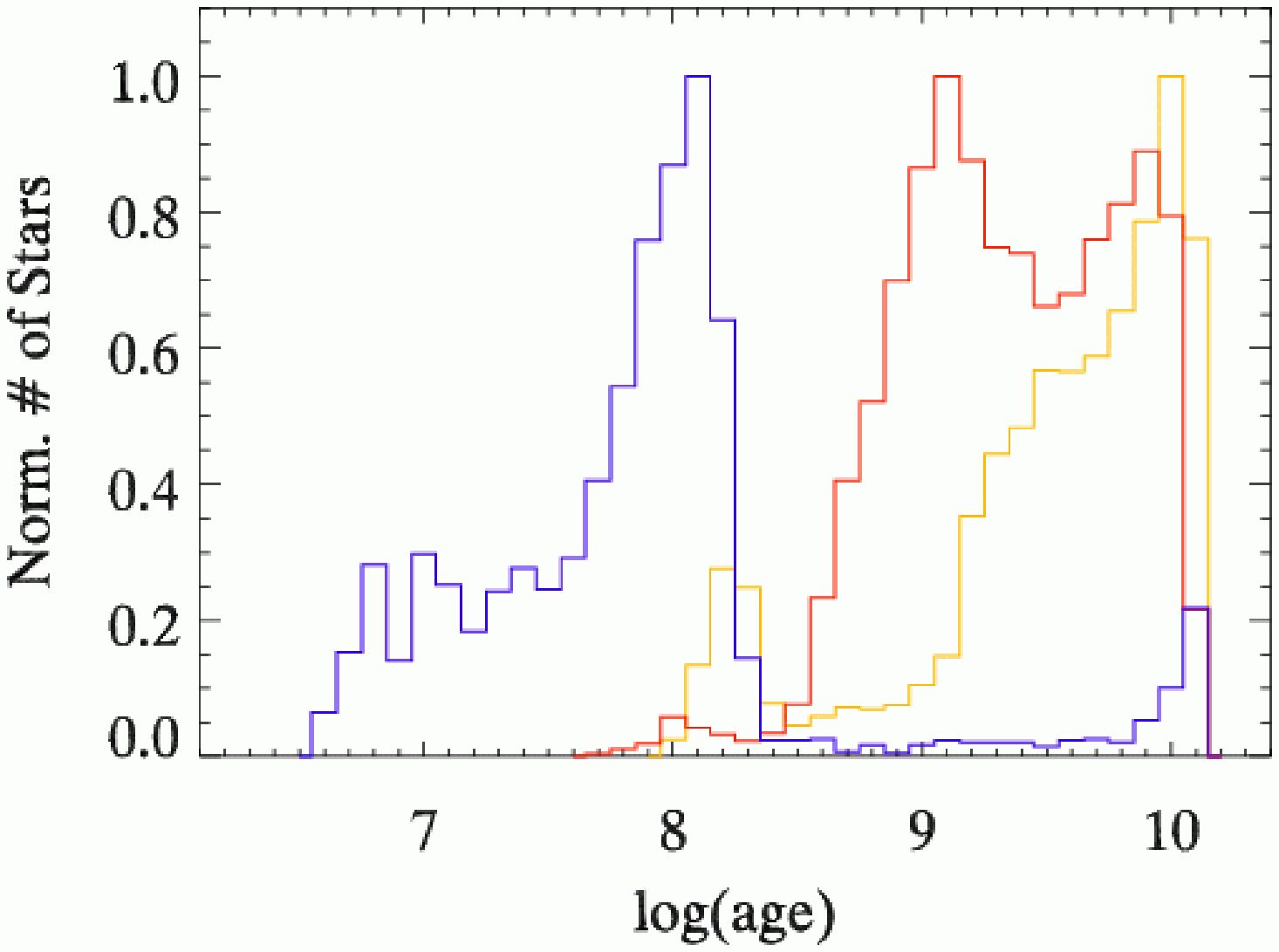}{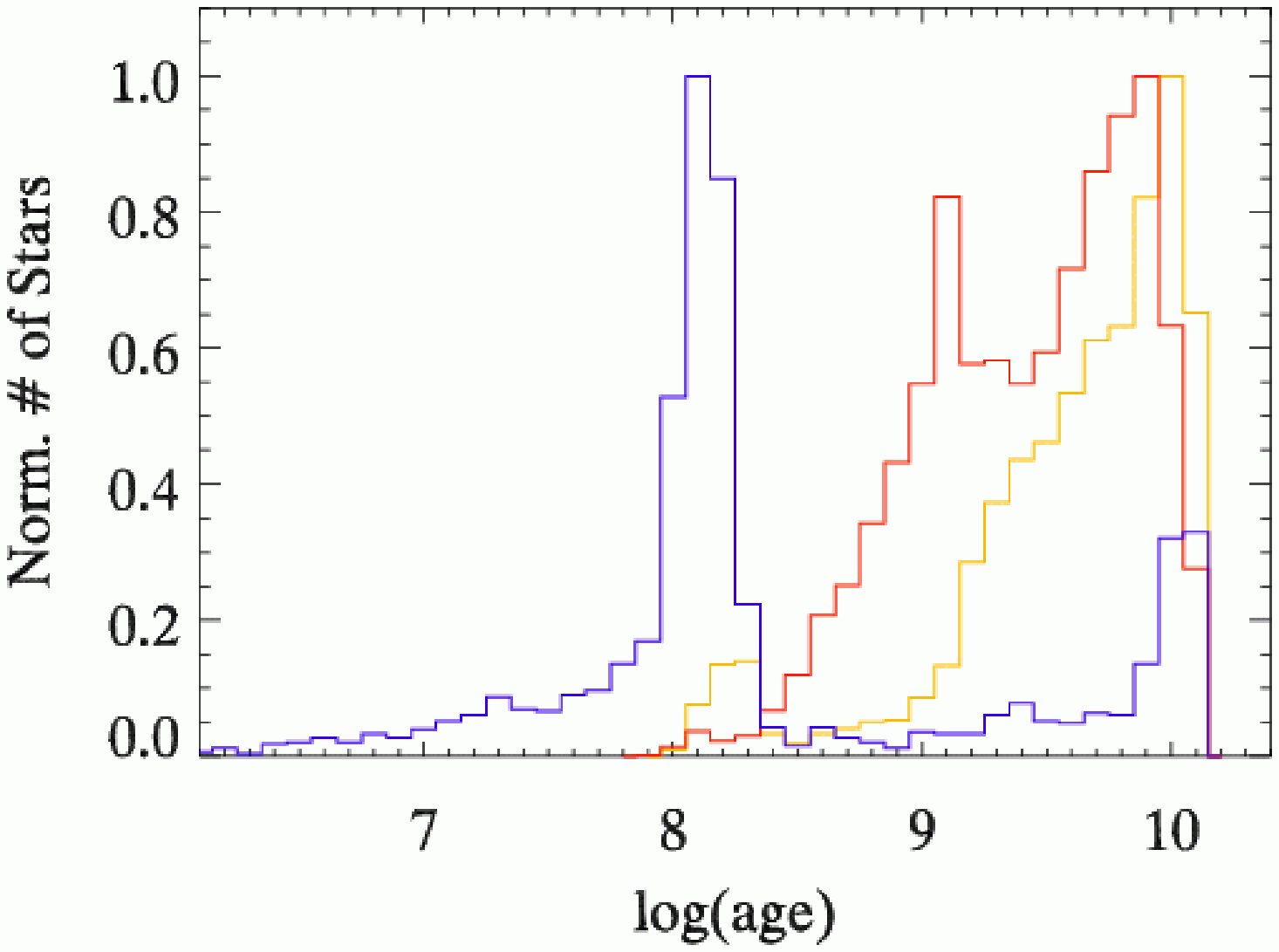}

\caption{Histogram of ages detected in defined CMD boxes (see Fig.~1)
  for NGC~4144 assuming a constant star formation rate from 13 Gyr to
  the present.  Histograms are based on synthesized CMDs created with
  the MATCH (left) and IAC-STAR (right) programs as described in
  \S4.1.  These plot shows that stars in the MS box (blue) are
  dominated by stars $\sim$100 Myr in age, while the AGB (red) and RGB
  (orange/yellow) boxes have typical ages of $\sim$1 Gyr and $\sim$10 Gyr
  respectively.  Similar plots for other galaxies are qualitatively
  similar.}

\end{figure*} 

\subsection{Stellar Density Profiles}

\begin{figure*}
\plotone{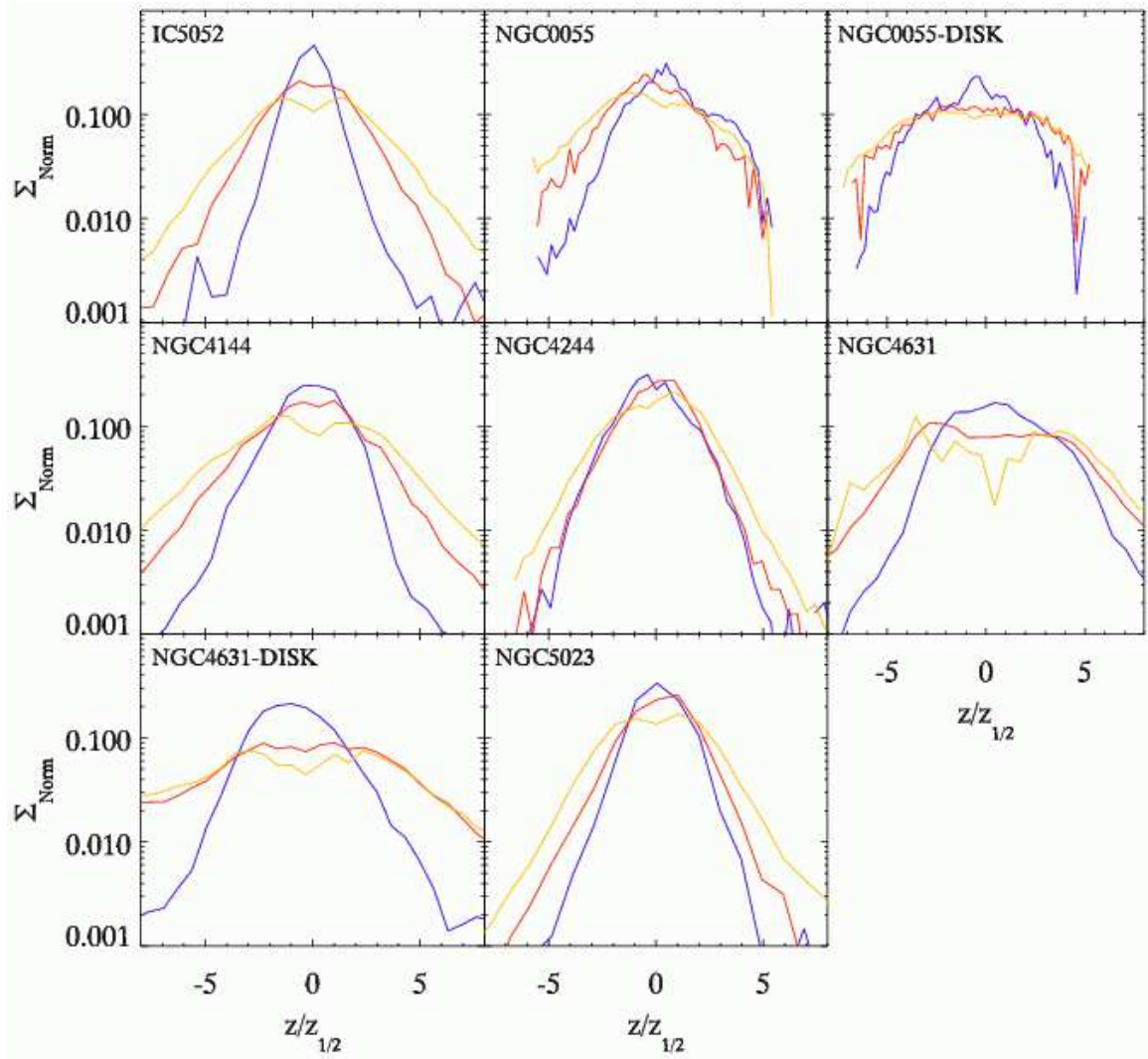}
\caption{The normalized surface density as a function of scaleheight
  for young MS (blue), intermediate-age AGB (red) and old RGB (orange/yellow)
  stars.  Each surface 
  density distribution was normalized to integrate to one.  Note that
  in all cases the MS distribution is the most peaked while the RGB
  distribution is the widest.}
\end{figure*}

Now we compare the surface density profiles of the MS, AGB and RGB
stars to examine possible variations in stellar population with disk
height.  Figure~5 shows the completeness-corrected profiles as a
function of disk height for each field.  Each profile is derived using
the same methodology as in \S3, typically using $\sim$10,000 stars per
field.  The surface densities are then normalized to have
$\int\Sigma dz = 1$.  All the fields show a similar pattern.  The
MS (blue) stars have the narrowest distribution while the AGB
(red) and RGB (orange/yellow) stars have broader distributions and typically
show a dip near the midplane.  Because we have corrected for
incompleteness, the dip almost certainly due to dust absorption, as we
demonstrate in \S4.4 with a very simple model.

Figure~5 suggests that older stellar populations become more prominent
with increasing disk height.  We quantify this trend in Figure~6, which
shows the ratios of surface densities in our different age bins.  The
ratios were normalized to those expected for a constant SFR using
the MATCH synthetic CMDs (see \S4.1.1).  A ratio of one in Fig.~6
therefore corresponds to a constant star formation rate and increasing
values correspond to older stellar populations.  We note
that the ratio is only plotted where the signal-to-noise of the ratio
is greater than 3.  The small number of MS stars at large scale
heights limits our ability to trace the $\Sigma_{\rm
RGB}$/$\Sigma_{\rm MS}$ and $\Sigma_{\rm AGB}$/$\Sigma_{\rm MS}$ as
high above the midplane as the profiles shown in Figures~3, 5 and 7.
Also, NGC~4631 and NGC~4631-DISK are not included in Figure~6 because
the high completeness limit results in very few  RGB stars (see Fig.~1) and
an increased contamination of AGB stars in the RGB box.

The top and middle panels of Figure~6 show that in each of the fields,
RGB stars become more numerous relative to MS and AGB stars with
increasing disk height.  However, this trend shows an enormous
variation from galaxy to galaxy.  In IC~5052 the ratio $\Sigma_{\rm
RGB}$/$\Sigma_{\rm MS}$ becomes as high as $\sim$100 times the
midplane value, while in NGC~55 and NGC~4244 the increase is much more
moderate, to $\lesssim$10 times the midplane value.  This variation is
most likely the result of a range of recent SFRs in our galaxies.  The
ratio $\Sigma_{\rm RGB}$/$\Sigma_{\rm AGB}$ is much more consistent
from galaxy-to-galaxy, however.  This may result from the overlapping
time range spanned by stars in the AGB and RGB boxes and/or the large
time ranges these boxes span relative to the MS.  We argue in \S5 that
the RGB population is likely to be dominated by truly old stars much
older than the AGB population.  Interestingly, the field showing the
the flattest $\Sigma_{\rm RGB}$/$\Sigma_{\rm AGB}$ profile was the
NGC~55-DISK field located in the outer parts of the NGC~55 disk,
perhaps suggesting a different star formation or dynamical history at
large radii ($\sim 5 \times h_{R}$).  However, this galaxy is the least
inclined in our sample and is somewhat irregular in shape, therefore
results for this one system should not be overinterpreted.

The low values of $\Sigma_{\rm AGB}$/$\Sigma_{\rm MS}$ and the high
values of $\Sigma_{\rm RGB}$/$\Sigma_{\rm AGB}$ result from a lack of
AGB stars compared to the constant SFR MATCH synthetic CMD.  This
would seem to suggest that the galaxies' star formation histories
(SFHs) are depressed at intermediate ages and enhanced at young ages.
However, as we noted in \S4.1, the AGB morphologies in the synthetic
CMDs are not well matched to the observational CMDs, probably because
of the difficulty in modeling the AGB phase of evolution
\citep{marigo01}.  This discrepancy combined with the differences seen
between the two sets of synthetic CMDs suggests that a derivation of
accurate SFHs using just the brightest stars in a galaxy is not yet
possible.

To check if the scatter in the $\Sigma_{\rm RGB}$/$\Sigma_{\rm MS}$
ratio was in part due to varying radial coverages of the galaxies
(Table~3), we remade the plots in Figure~6 using only stars within the
central scale length of each galaxy.  These plots were similar to
those shown and showed comparable scatter.  This suggests that the
observed variations from galaxy to galaxy in the stellar populations
ratio reflect global differences in the galaxies' SFHs and/or vertical
structure.  For instance, if we assume the trend towards older
populations with increasing scale height results from disk heating,
then the scatter in Figure~6 suggests substantial variations between
galaxies in either the mechanisms that heated the disk, or the SFH of
the disk.  Despite these variations, Figure~6 gives strong evidence
that overall the age of the stellar populations increases with
increasing scale height.

\begin{figure}
\plotone{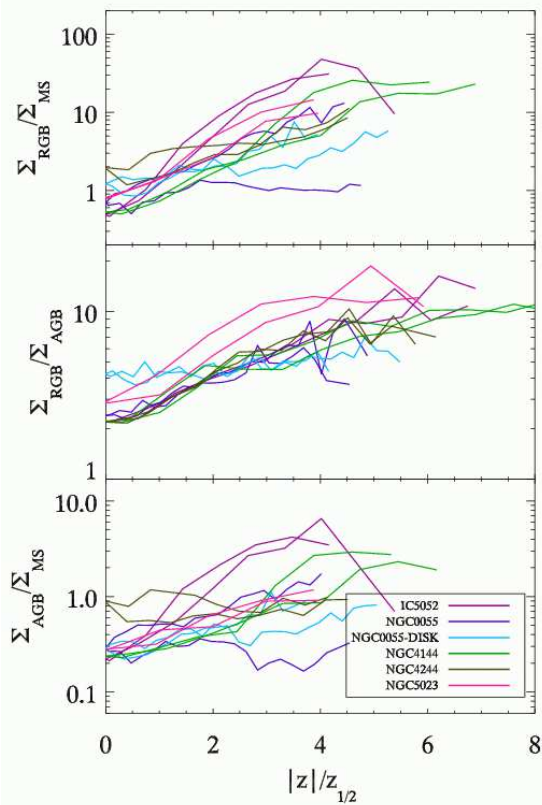}
\caption{The ratio of the surface density of different stellar
  populations as a function of disk height:  RGB/MS (top), RGB/AGB
  (middle), and AGB/MS (bottom).  Points were normalized based on
  the number counts of a constant star formation rate model, e.g. a
  value of 1 suggests a constant star formation rate, while values
  $>$1 suggest a decrease in the star formation rate with time.  Note
  that the MATCH synthetic CMDs were used in this normalization
  (\S4.1.1).  Only points with a 
  signal-to-noise $>$ 3 are plotted.  The plots clearly show that the
  stellar populations become older with increasing disk height.}  
\end{figure}

\subsubsection{Stellar Population Scale Heights}

To further quantify the differences in the vertical distribution of
the three CMD regions, we fit each surface density profile to a
sech$^{2}$ function in which the normalization, central position,
scale height ($z_0$), and background level were all allowed to vary.
We fit each profile only at disk heights $>$3$z_{1/2}$ to avoid the
dips near the midplane, except in IC~5052 and NGC~5023 where we used
disk heights $>$1.5$z_{1/2}$ to allow fitting of the very narrow
$\Sigma_{\rm MS}$ profiles.  Figure~7 shows the resulting fits to the
$\Sigma_{\rm MS}$, $\Sigma_{\rm AGB}$, and $\Sigma_{\rm RGB}$ profile
of each galaxy, in the top, middle and bottom panels respectively.
The observed profiles are shown as a solid line, while the best
fitting sech$^2$ function is shown as a dashed line.  The dotted lines
at $\pm$3$z_{1/2}$ ($\pm$1.5$z_{1/2}$ in IC~5052 and NGC~5023) delineate
the region excluded from the fit.  The error bars on the
data points are used to weight the sech$^2$ fits and reflect Poisson
errors in the number counts, but do not include uncertainties in the
completeness corrections.  The scale height of the best-fitting
sech$^2$ function is shown in the upper-left corner of each panel, and
the error shown is scaled by the square root of the reduced $\chi^2$
of the fit.  The reduced $\chi^2$ values for a majority of the fits
were between 0.8 and 1.3, but were larger for NGC~55 and NGC~4631
due to their irregular structure.  Scale heights and errors for all
the fits are shown in Table~4.

We find that in each galaxy, the MS scale height value is the
narrowest followed by the AGB and then the RGB.  In all cases the RGB
population is significantly broader than the AGB, MS and $K_s$ band
$z_0$ values.  This result strongly suggests the presence of an older
component with larger scale height.  An analysis of the variations in
scale height of the MS, AGB and RGB populations with scale length in
each of the galaxies turned up no obvious trends.  We also identified
no trends with galaxy rotation speed, due to the small range of masses
spanned by our sample galaxies.  We note that in some cases, the $z_0$
derived for all stars (\S3) is somewhat larger than the $z_0$ derived
for just the RGB stars.  This results from the lower scale heights
used in the fits to the different stellar populations - if the lower
limit for the fit to all stars is reduced from 5$z_{1/2}$ to
3$z_{1/2}$, the derived $z_0$ is less than or equal to the RGB star
$z_0$ in each galaxy, as expected if the total stellar density is well
characterized by a combination of the MS, AGB and RGB components.  The
fits presented here for these different stellar population components
are in general not extremely sensitive to the range of $z$ values
used.  Varying the lower limit of the fit between 1.5-5$z_{1/2}$
typically changed the AGB and RGB $z_0$ values by less than 10\%.  The
compact MS components were more dramatically affected because of the
smaller number of stars at large scale heights.

From this analysis it appears that NGC~4631 has the ``thickest'' old
component with an RGB scale height of $\sim$1250~pc, roughly 2.5 times larger
than the MS and $K_s$ band fits.  We note that the fits for
NGC~4631 were truncated at large negative disk heights to prevent the
contamination of stars from companion galaxy NGC~4627.  IC~5052 and
NGC~4144 have similar ratios ($\sim$2.5) of RGB to MS scale heights, while
NGC~4244 has the smallest ratio, with an RGB scale height only 1.7 times 
that of the MS stars.

There is also evidence for a modest flaring of the stellar components
between the central and '-DISK' pointings of  NGC~55 and NGC~4631.
The '-DISK' fields are centered 4.8 and 6.1 scale lengths (see
Table~2) from the center of the NGC~55 and NGC~4631 respectively.
An increase in $z_0$ values by a factor of 1.1 to 1.6 is seen for all
three components in NGC~55 and for the AGB and RGB component in
NGC~4631.  

The profiles deviate from the fitted sech$^2$ profile significantly
near the midplane.  This deviation is almost certainly due to dust, we
model this effect in \S4.4.  At larger disk heights ($>$2-3~kpc) there
is also a slight overdensity in the RGB components of IC~5052,
NGC~4144, NGC~4244 and NGC~5023.  These overdensities hint at the
possible presence of an even more broadly distributed old component.
In three of these galaxies, the RGB fits had elevated $\chi^2$ values
relative to the MS and AGB fits.  We estimate that these overdensities
occur at a surface brightness $\mu_{\rm F606W} \gtrsim$ 28 mag
arcsec$^{-2}$ assuming a luminosity function similar to galactic
globular clusters \citep{buonanno94,kravtsov97}. However, without
better knowledge of the background level, it is not
possible to verify the existence of this component.

The scale heights measured in the $K_s$ band (Table~4) are closest to
those measured for the MS and AGB components.  Half of the galaxies
have $K_s$ band scale heights closer to the MS value and the other half
closer to the AGB value.  This suggests that the $K_s$ band light in
these galaxies is dominated by relatively young stellar populations,
probably red supergiants and AGB stars \citep[in agreement with the
  findings of][]{aoki91}.  This result runs contrary to
the common assumption that the NIR light primarily traces older stellar
populations \citep[e.g.][]{florido01}, and is significant in that NIR
luminosity is often used as a proxy for stellar mass when comparing
galaxies of different types and masses.  However, we note that our
$K_s$ band scale heights are biased towards higher surface brightness
populations due to the bright limiting isophote of the 2MASS data from
which they are derived.  

\begin{figure*}
\plottwo{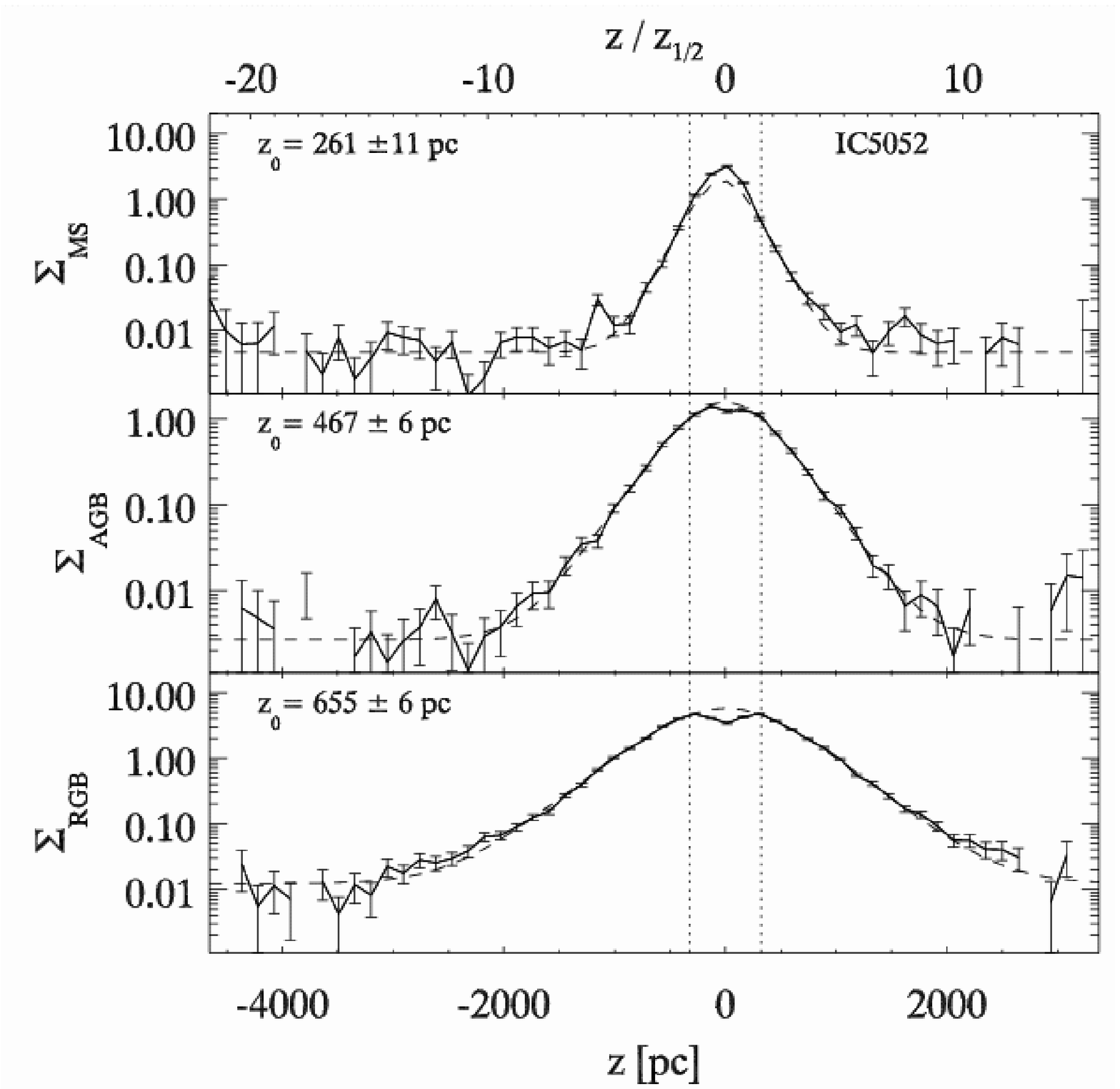}{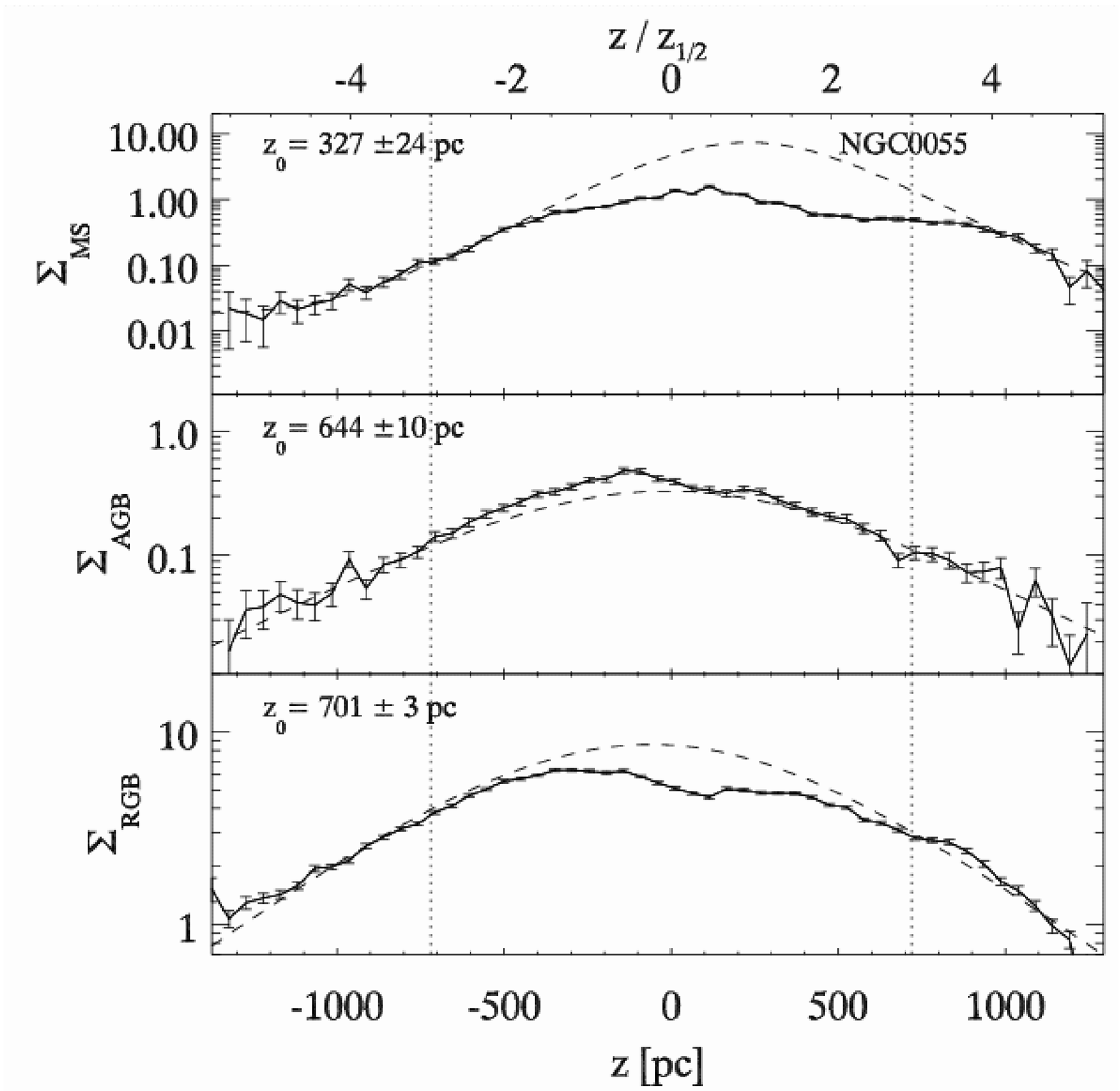}
\plottwo{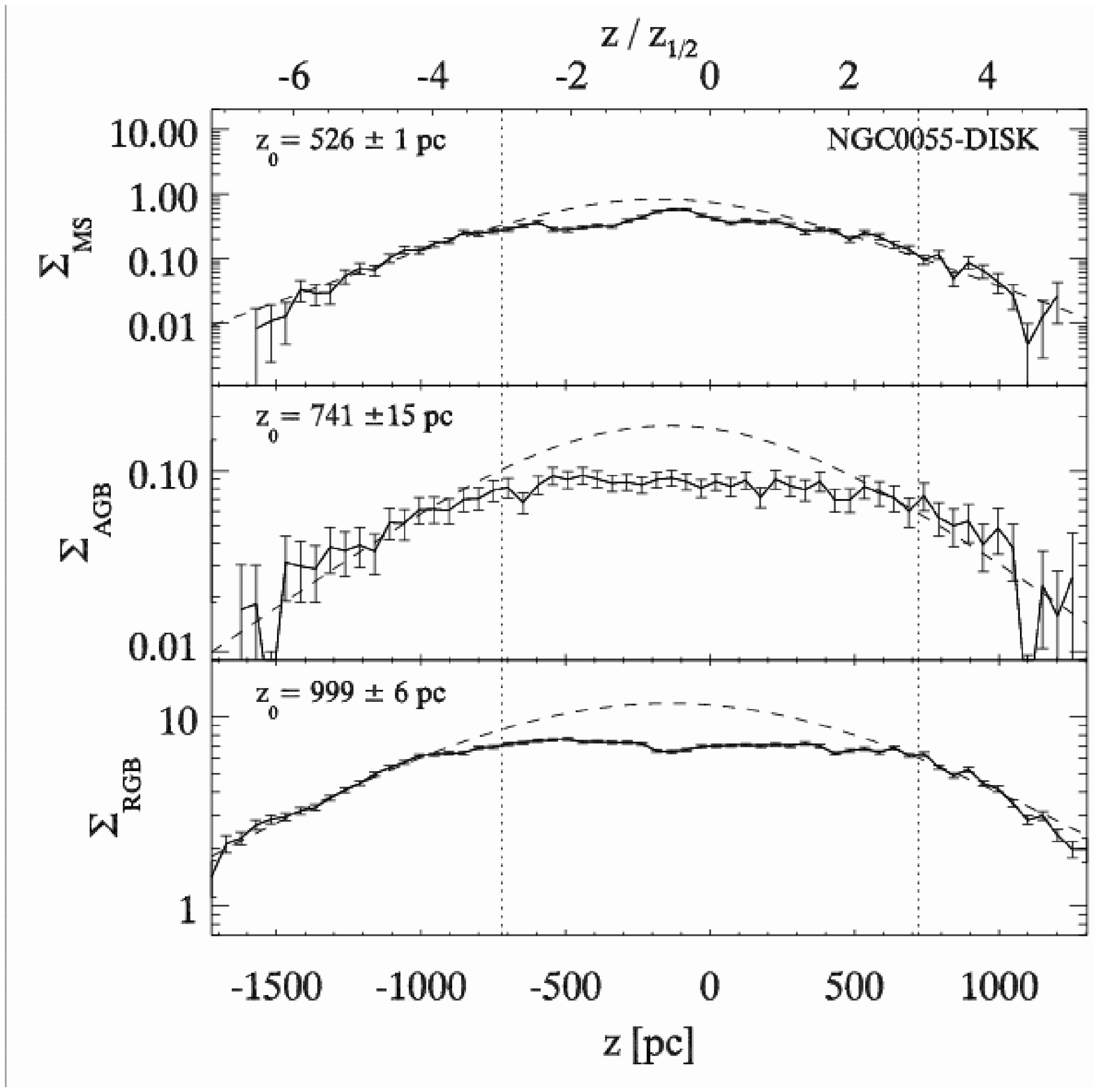}{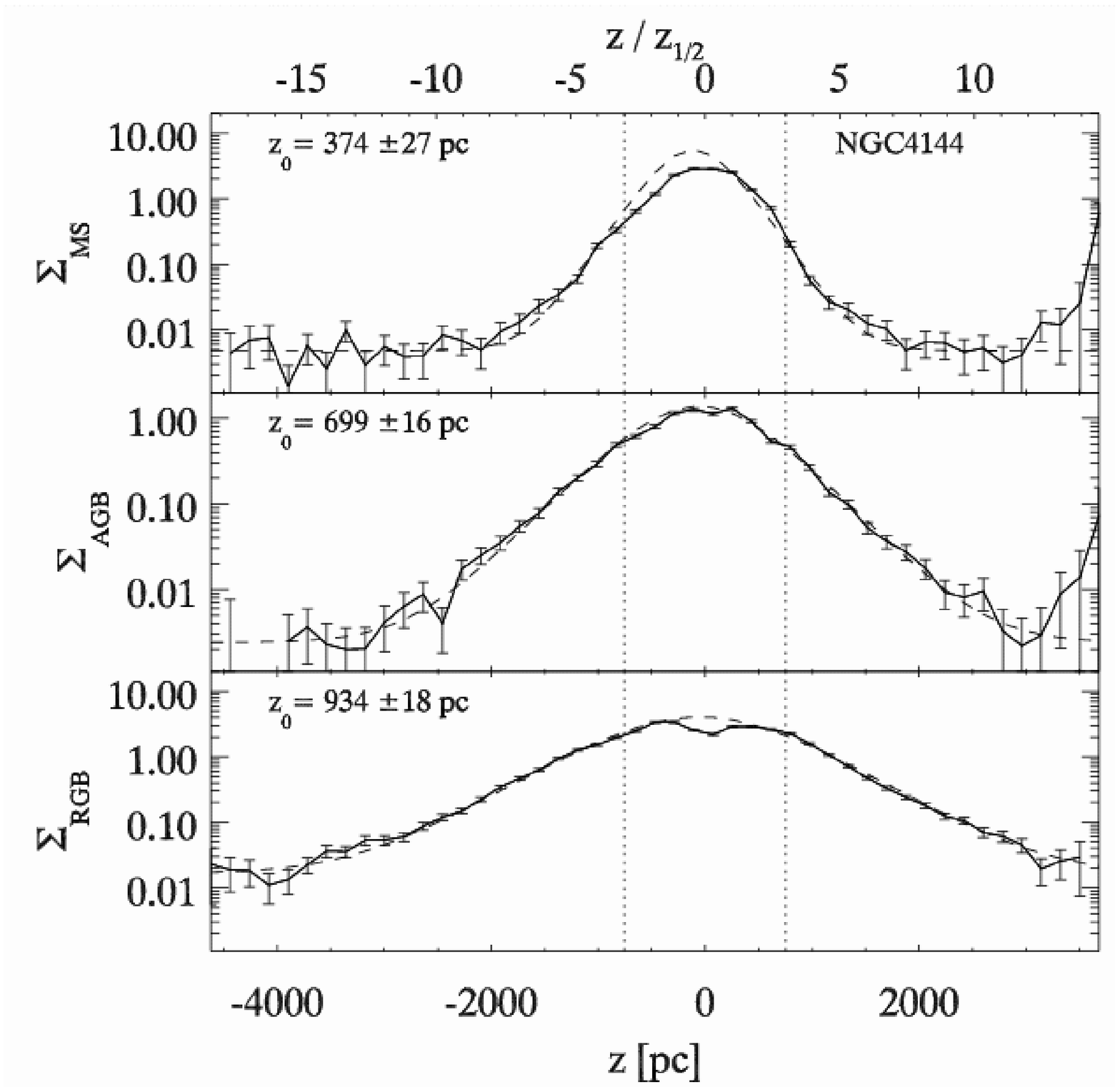}
\caption{Model fits to the observed surface density distribution as a
  function of disk height in each galaxy for the MS (top), AGB
  (middle), and RGB (bottom).  The dashed line shows the best fitting
  sech$^2$ model excluding data with $|$z/z$_{1/2}|<3$ (as shown with
  the dotted lines) while the vertical dotted lines show the range of
  $z$ values excluded from the fit.  The number in the upper left
  corner gives the scale height of the best fitting sech$^2$
  function.}
\end{figure*}

\addtocounter{figure}{-1}
\begin{figure*}
\plottwo{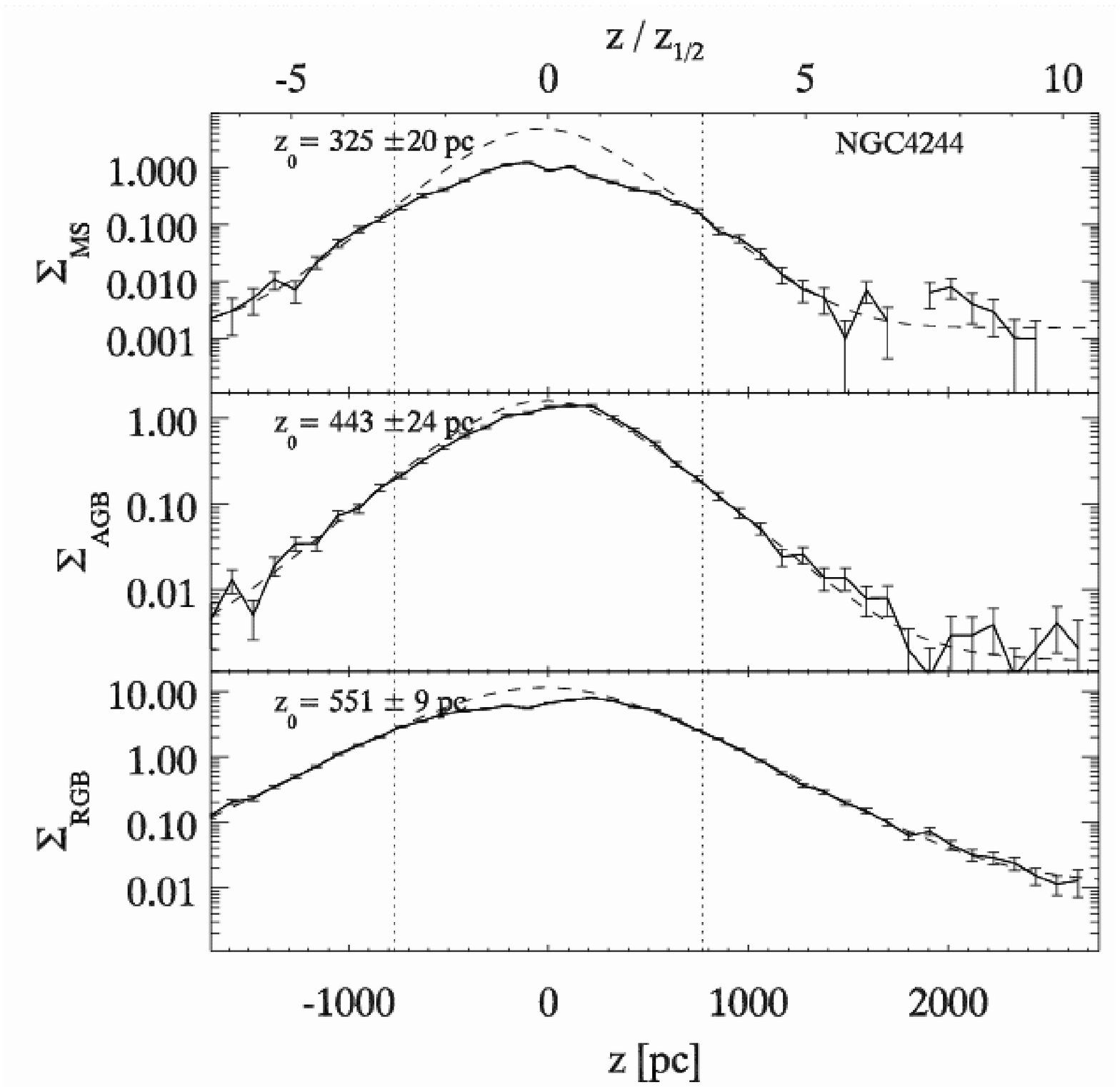}{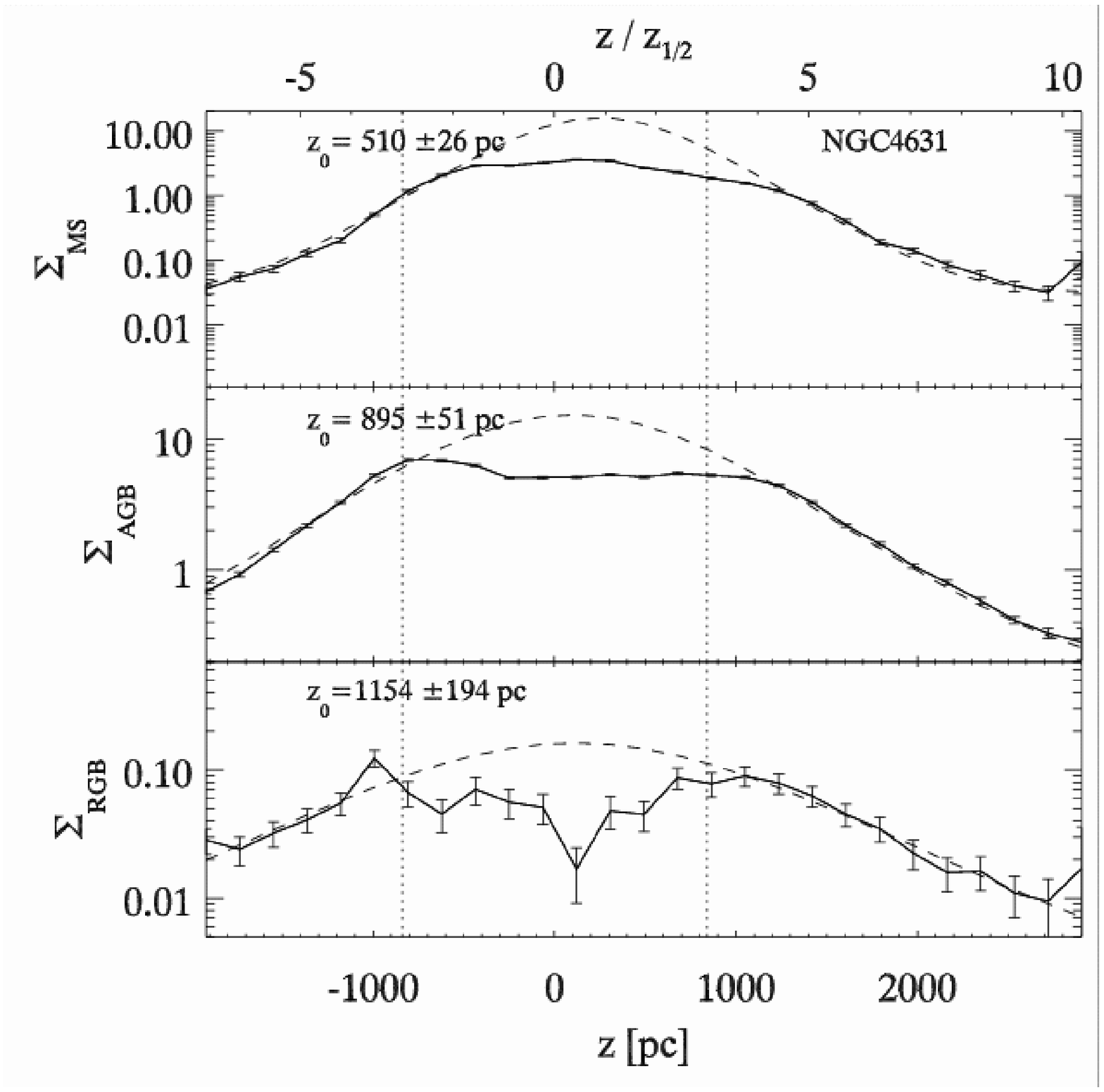}
\plottwo{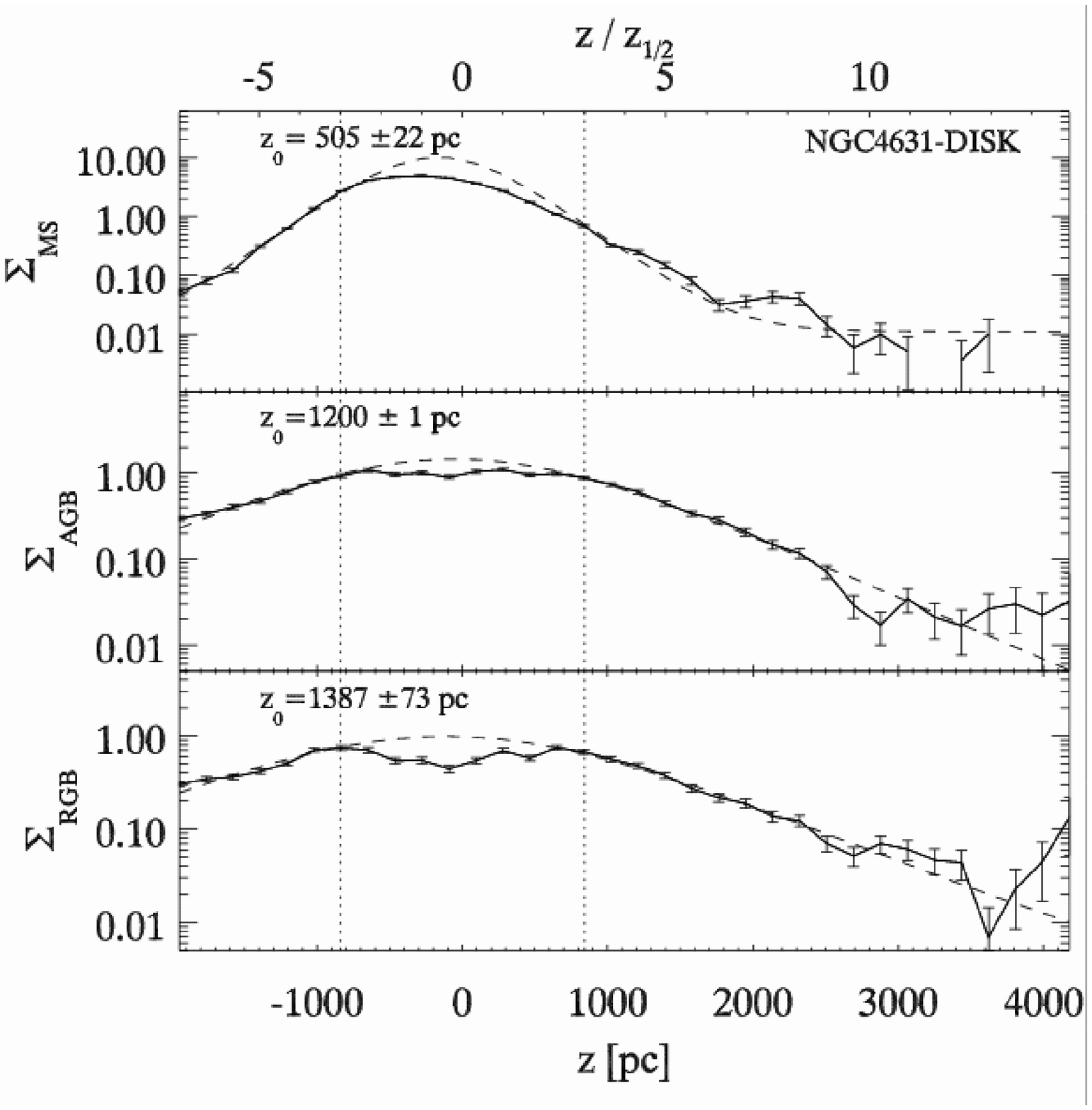}{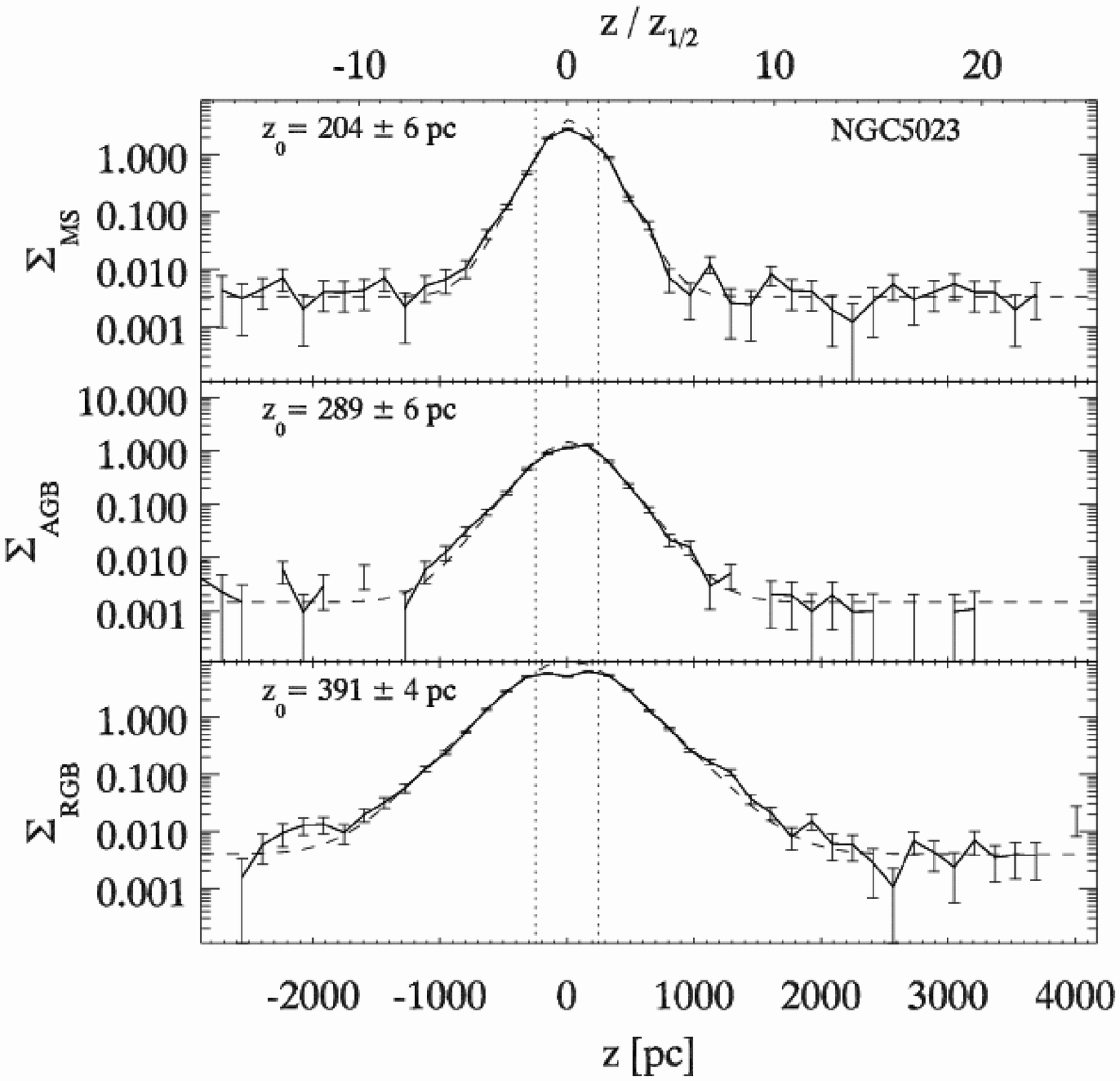}
\caption{{\it continued}}
\end{figure*}

\subsubsection{Comparison to Previous Observations}

The results above indicate that there is a systematic increase in the
vertical scale heights of older stellar populations in our sample of
low mass, late-type disks.  Before investigating possible origins for
these structural differences in
\S\ref{diskheatingsec}~\&~\S\ref{metallicitysec}, we now compare our
measurements of scale heights to previous observations of the vertical
structure of disks.

The most detailed constraints on the scale heights of different
stellar populations come from the solar circle of the Milky Way.
Studies have revealed a complicated disk structure, with a young and
old thin disk embedded within a more extended thick disk.  The young
thin disk is the narrowest of the three, having a scale height of 
z$_0 \sim 200$~pc, as traced by stars with bright absolute magnitudes 
($M_V \lesssim 3$) \citep{schmidt63}.  In contrast, the scale heights of the
young main sequence stars in our sample are almost all significantly
larger than the Milky Way value suggesting that the low mass galaxies
in our sample form stars in a thicker layer than the Milky Way,
consistent with \citet{dalcanton04}.  The resulting axial ratios for
our samples' young star forming disks are also much thicker as well,
with $z_0/h_r=1.8-6.3$ (see Table~4) for our sample galaxies, versus
$z_0/h_r \sim 15$ for the young thin disk of the Milky Way.

The division of the Milky Way's older stellar populations into a thin
and thick disk was first introduced by \citet{gilmore83} to explain a
break in the number counts of F \& G stars at $\sim\!1$~kpc.  While
the need for two old disk components was long debated, recent
measurements of systematic $\alpha$-element enhancement in thick disk
stars \citep[most recently][]{gratton03,feltzing03,mishenina04,bensby05} 
strongly suggest that the thick disk is indeed distinct from the old
thin disk.  Recent observations \citep{chen01,siegel02} give a scale
height for the old thin disk of z$_{0}\!\sim\!600$~pc, similar to
found in the original \citet{gilmore83} study.  These same studies
suggest that the exponential scale height of the thick disk is
$h_z\!\sim\!700$~pc (corresponding to z$_0 \sim 1400$~pc, thinner than
originally claimed).  The Milky Way thick disk is therefore roughly twice
the height of the old thin disk, and 7 times the height of the young
thin disk.

Within our own sample, the scale height of the old RGB component is
mostly intermediate between the Milky Way old thin disk and the thick
disk.  Our sample galaxies have much lower masses and surface
densities than the Milky Way, and, lacking any firm model that
predicts how the properties of the old thin disk and the thick disk
should vary with galaxy mass, we are hesistant to attribute the
extraplanar population to either an old thin or a thick disk on the
basis of the the surface brightness profiles alone. There are no
dramatic inflection points in the RGB surface density profiles plotted
in Figure~7 that would assist in a unique separation of old thin disk
and thick disk stars, and the possible overdensity of stars above
2-3~kpc may well be due to a stellar halo.  Even if the RGB component
is similar to the Milky Way thick disk, this lack of inflection is not
unexpected.  In the Milky Way, the inflection point in the
surface density of F \& G dwarves that marks the separation of the
thin and thick disks is likely the result of two different
populations of stars separated in age.  The lack of similar inflection
points in our RGB profiles can easily be explained if the RGB stars
don't have as wide a range of ages as the Milky Way dwarves.  We will
show in \S5 that the RGB stars in our galaxies may very well be
dominated by a single-age population.

We do note, however, that the axial ratios of the RGB disks range
between $h_r/z_0=1.0-3.3$, with a median of 1.8 (adopting the
$K_s$-band radial scale length, and averaging the two independent
measurements for NGC~55 and NGC~4631).  For comparison, the axial
ratios of the old thin and the thick disks of the Milky Way are 5.0
and 2.1, respectively (assuming $h_r=3$~kpc for both components).
Thus, in terms of their {\emph{overall}} structure, the RGB component
we detect is significantly more round than the Milky Way's old thin
disk, and is distributed more like the Milky Way thick disk.  However,
without additional information we cannot ascribe a common formation
scenario to our observed RGB component and the Milky Way thick disk.
We revisit this issue in the discussion (\S\ref{discussionsec}), after
analyzing the disk heating and the vertical metallicity gradients of
our sample galaxies.

Outside of the Milky Way, the most detailed information comes from
studies of the vertical distribution of resolved stars in HST images,
similar to the work we present in this paper.  \citet{tikhonov05a} and
\citet{tikhonov05b} present evidence for extended components in six
galaxies, which they qualitatively argue correspond to thick disks and
halos.  Of the galaxies that overlap our sample (NGC~55, NGC~4144, and
NGC~4244), they include archival WFPC2 observations to reach greater
disk heights in NGC~4244 and NGC~55 than spanned by our ACS images.
In both cases they assume {\emph{a priori}} that the RGB stars at
lower disk heights trace a thick disk.  For NGC~4244,
\citet{tikhonov05b} show an exponential distribution of RGB stars
between $\sim$1 and $3$~kpc (their Figure~8) that appears to roughly
match the scale height of the profile shown in Fig.~7.  Beyond
$3$~kpc, they see a flattening in the number counts which they claim
is a halo, but which may also be the background level.  For NGC~55,
\citet{tikhonov05a} plot an exponential distribution of RGB stars
between 2 and $7$~kpc -- i.e. at much greater disk heights than probed
by our data.  However, based on inspection of their Figure~12, the
extended RGB component has a z$_0$ value of $\sim\!2$~kpc, which is
2-3 times the width of the RGB component we fit.  Although they assume
this component is due to a thick disk based on its exponential surface
density distribution, the axial ratio of this component would in fact
be prolate ($h_r/z_0\!\sim\!0.5$) and thus may be more analogous to
the Milky Way's stellar halo\footnote{Note, however, that NGC 55 is
the least inclined galaxy in our sample, complicating the
interpretation of its projected structural parameters.}.  The change
in slope also implies a break in the RGB distribution in NGC~55 at
around $2$~kpc.  By analogy, this may indicate that the marginal
overdensities we are seeing at comparably large disk heights in our
RGB profiles might be the signature of an additional broader halo
component.  \citet{mould05} also finds the presence of old stars at
large scale heights, and while these stars are automatically assumed
to be a thick disk component, no detailed analysis of their spatial
distribution is presented.

In addition to these recent studies of resolved stars, most previous
studies of the vertical structure of disks have focused on detecting
thick disks and stellar halos using unresolved surface brightness
profiles of the galaxies
\citep[e.g.][]{pohlen04,fry99,dalcanton02,neeser02}.  Because we only
detect stars at bright magnitudes, it is difficult to accurately
convert our measured surface density of stars (Figure~2) to a surface
brightness.  However, assuming the outer parts of our galaxies have
luminosity functions similar to Galactic globular clusters
\citep{buonanno94,kravtsov97}, we estimate that we reach F606W surface
brightnesses of $\sim$28 mag arcsec$^{-2}$.  This is comparable to the
depth reached in deep ground-based observations.

Only one of our galaxies has been analyzed for vertically extended
components using ground based data.  \citet{fry99} present $R$-band
surface photometry of NGC~4244 and find no evidence of a second thick
disk component, based on the lack of an inflection point in the
surface brightness distribution above the plane.  They trace the
vertical profile of NGC~4244 along the minor axis to $\sim\!2$~kpc at
which point it falls below their surface brightness limit of 27.5 mag
arcsec$^{-2}$.  We trace the RGB component out to nearly $3$~kpc, and
find a scale height that is similar to their fitted $R$-band
scale height (assuming $h_z$=$\frac{1}{2}$z$_0$).  Their lack of an
inflection in the surface brightness profile is consistent with the
\citet{tikhonov05b} analysis to larger scale heights.  This suggests
that their fit was dominated by the old stars, and not a younger
population.  However, the lack of an inflection in the surface density
distribution does not unambiguously rule out the presence of multiple
components.

The ubiquity of thick disks in galaxies has previously been proposed
by \citet{dalcanton02} based on color-gradients in edge-on disk
galaxies.  Our observations confirm that the color-gradients (at least at
the low-mass end) are the result of true differences in stellar
populations.  However, whether these gradients have an analagous
formation mechanism to the Milky Way thick disk is not clear.

One set of observations that reaches considerably deeper than these
ground-based observations is presented by \citet{zibetti04a}, who used
stacked Sloan images to show that halos are common in late-type,
edge-on galaxies.  Their composite galaxy has a significantly wider
field of view and poorer resolution than our observations.  They show
that the best-fitting model to their data is a disk+halo model, with
the disk component dominating out to roughly 10 exponential scale
heights ($\sim$10z$_{1/2}$).  Their limited resolution and combination
of a heterogenous sample of galaxies would likely prevent them from
seeing the RGB components we see in our galaxies.  However, the
possible detection of the more extended RGB components detected in
IC~5052, NGC~4144, and NGC~5023 may be halos similar to the
\citet{zibetti04a} halo.

\subsection{Disk Heating} \label{diskheatingsec}

The increase in $z_0$ seen in each galaxy between the MS, AGB and RGB
populations (Fig.~7, Table~4) could result from a number of
mechanisms, including vertical heating of a thin disk.  Such a model
would naturally produce the observed trend of older stellar
populations having larger scale heights.  In Figure~8 we plot the
increase in scale height with mean stellar age for four fields that
span the observed behaviour in our sample.  We plot scale heights for
the RGB and AGB, normalized by the MS scale heights, with the height
of the symbols indicating the 1$\sigma$ uncertainties on $z_0$.  Note
that when interpreting our data in the context of disk heating models
we are therefore implicitly assuming that the RGB and AGB stars were
originally formed in a layer with a scale height comparable to that of the
present day main sequence stars.  We assign characteristic ages to the
RGB and AGB using the MATCH synthetic CMD tests for a constant
SFR (\S4.1.1).  However, because the galaxies' actual star formation
histories may differ significantly from the constant SFR assumed in
Figure~4, we cannot assign a single age to each stellar population.
Instead, we use the resulting age distributions to identify the 25th,
50th (Median) and 75th percentile ages.  The resulting age ranges are
shown by the width of the individual boxes in Figure~8.  However,
note that the actual age of the population may lie entirely outside of
the boxes, for example, if the RGB stars were all formed in a single
burst 12~Gyr ago.  Thus, when interpeting Figure~8, one has substantial
allowance in assigning an age.

Overplotted on Figure~8 are dashed lines showing a range of power-law
increases in the disk scale height $z_0$ with time ($z_{0} \propto
t^{-\beta}$).  For an isothermal sech$^2$ profile, the $z_0$ values
are related to the vertical velocity dispersion ($\sigma_z$):
\begin{equation}
z_{0} = \frac{\sigma_{z}^{2}}{2 \pi G \Sigma} 
\end{equation}
where $\Sigma$ is the surface density of the disk \citep[Eq. 17
in][]{vanderkruit88}.  Studies of disk heating traditionally use power
laws in the velocity dispersion, $\sigma_{z} \propto t^{-\alpha}$, and
thus $\alpha = \beta/2$.  Figure~8 therefore demonstrates that the
vertical velocity dispersion of our galaxies has increased no faster
than $\alpha = 0.15$.  More specifically, there are no characteristic
ages that can be assigned to the AGB and RGB stars that yield heating
rates greater than $\alpha = 0.15$ (with the possible exception of the
more massive, interacting galaxy NGC~4631), and thus this conclusion
is robust even in light of our substantial age uncertainties.

%\begin{equation}
%\rho_{0} = \frac{\Sigma}{2\sqrt{2}z_{0}}
%\end{equation}
%Where $\Sigma$ is the surface density.

In contrast, the disk heating that has been observed in the Milky Way
is comparatively rapid.  The age-velocity dispersion relation (AVR)
for Milky Way disk stars suggests that the vertical velocity
dispersion increases with time with values of $\alpha$ ranging between
0.3 and 0.6 \citep[e.g.][see summary in Table 1 of Hanninen \& Flynn
2002]{wielen77,binney00,nordstrom04}.  In contrast, our limit of
$\alpha \lesssim 0.15$ is significantly smaller than the Milky Way
value.  These data immediately suggest that any disk heating in our
low mass galaxies has been far less effective than in the Milky Way.
Moreover, if some fraction of the extraplanar RGB stars are not due to
disk heating, and are instead due to direct accretion or {\emph{in
situ}} formation at large scale heights, or if the RGB stars are
weighted towards old ages (as we argue below in \S5), then the actual
rate of disk heating is even lower than suggested by Figure~8.

There are several reasons why disk heating is expected to be low for
our sample galaxies.  Within the Milky Way, the increase in vertical
velocity dispersion with time is thought to be due to scattering by
spiral arms \citep{barbanis67,sellwood84,carlberg85}, by molecular
clouds \citep{spitzer51}, or both \citep[][see also the review by
Lacey 1991]{carlberg87,jenkins90,jenkins92,shapiro03}.  However,
our galaxies have sufficiently low masses and surface densities that
they are unlikely to be globally gravitationally unstable
\citep{dalcanton04,verde02} and thus would not host strong spiral
arms.  Given that scattering by spiral arms seems to be the dominant
heating mechanism in the Milky Way \citep[e.g. most
recently][]{desimone04}, the absence of spiral arms alone should cause
a drastic drop in heating rate down to $\alpha\sim0.2-0.25$, the
expected value for heating by giant molecular clouds alone
\citep[e.g.][]{hanninen02}.  Likewise, the absence of strong dust
lanes in these systems and the results of \S\ref{dustmodelsec} both
indicate that the cold molecular ISM is in a thicker layer than in the
Milky Way.  This large scale height for the cold ISM, and the general
lack of molecular gas in low mass galaxies \citep{young91,leroy05}
should therefore further suppress the efficiency of disk heating.
Finally, the young stellar disks in our sample are apparently much
thicker than in the Milky Way, which could reduce the efficiency of
any heating mechanism \citep{freeman91}.  \citet{shapiro03} also argue
for reduced disk heating in late-type galaxies based on the ratio of
vertical to radial velocity dispersions.  However, their rationale for
the observed trend is opposite from what we conclude from our data.

As an aside, the low observed heating rates may provide strong
constraints on cosmologically important sources of disk heating
including late-time satellite accretion \citep{quinn93}, massive black
holes \citep{lacey85}, or halo substructure
\citep[e.g.][]{hanninen02,benson04}.  However, the expected heating
rates for such models have been calibrated for massive spiral disks,
not the thicker, lower surface density galaxies studied here.  

\begin{figure}
\plotone{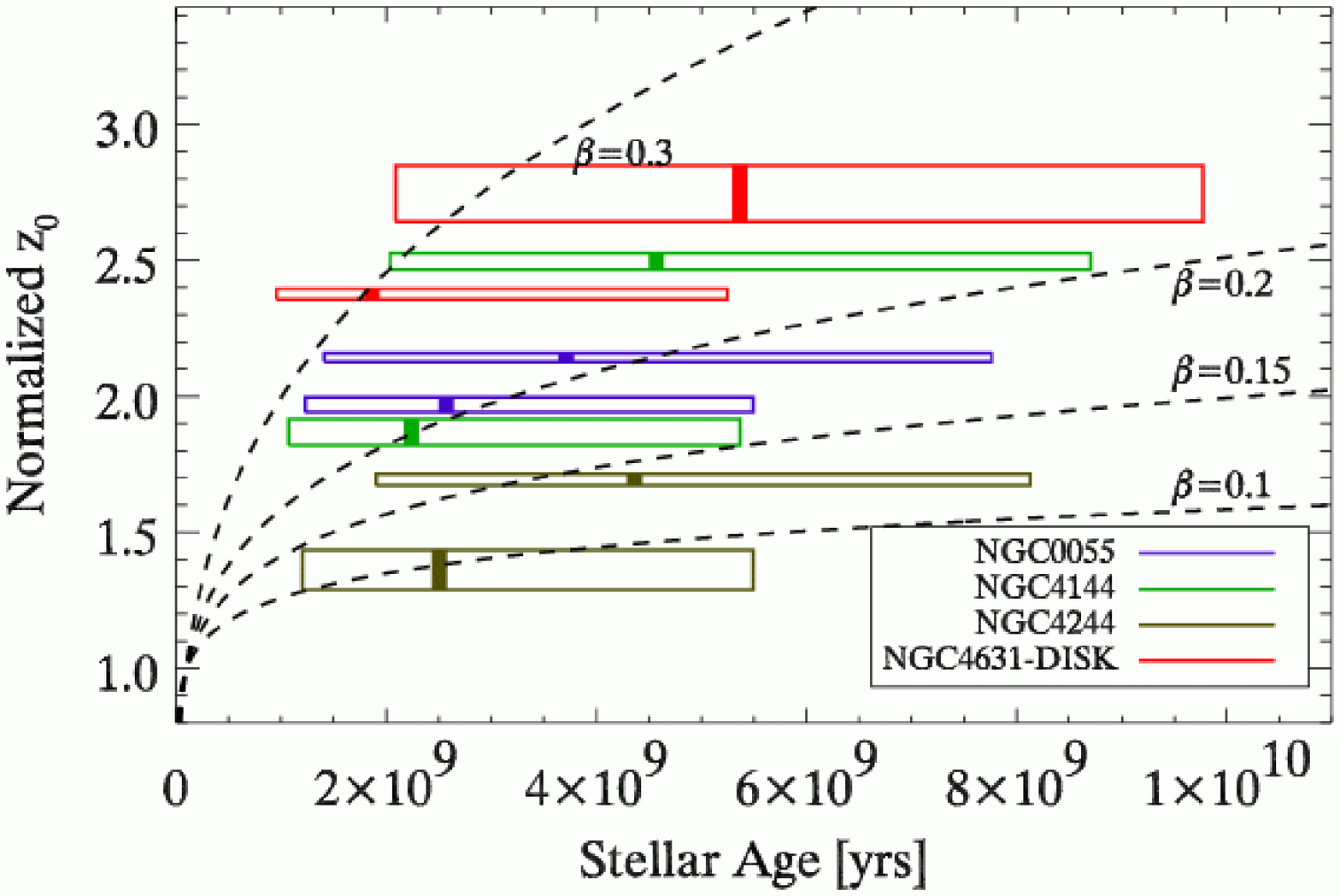}
\caption{This figure shows the increase of scale height with time
  based on the stellar populations analysis described in \S4.1-2. The
  AGB and RGB scale heights are shown for four galaxies spanning the
  range of $z_{0}$.  Each scale height is shown with a box, color
  coded by galaxy. The box width represents the 25th, 50th, and 75th
  percentile age range of the AGB or RGB population (see Fig.~4) and
  the height represents the 1$\sigma$ errors in the determined $z_0$
  (Fig.~7).  The $z_0$ values were normalized by the $z_0$ of the MS
  population.  The dashed lines show power-laws of the form $z_{0}
  \propto t^{-\beta}$.  The RGB (upper) bar and AGB (lower) bar for
  any galaxy can be used to roughly constrain the disk heating rates.
  Because the precise star formation history of these components is
  not known, the width of the boxes is derived from a constant star
  formation rate model and thus represents the range of ages that
  contribute to the AGB and RGB population.}
\end{figure}

\subsection{Modelling Dust Effects on the Stellar Density Profiles} \label{dustmodelsec}

Before continuing to explore the origin of extraplanar stars, we
briefly examine the vertical distribution of the dust layer.  At first
glance, interpretation of the stellar density profiles near the
midplane in Figure~5 might be somewhat confusing.  If the dips in
surface density are due to dust, why does the dust appear to affect
the RGB and AGB stars more than the MS stars?  We suggest this may
occur because the dust layer is opaque near the midplane and is
distributed with a scale height greater than or equal to the MS
population, but less than the AGB/RGB populations.  The MS stars we
are seeing would then lie entirely in front of an obscuring dust
screen, while the AGB/RGB populations would have a significant
population at large disk heights above where the galaxy becomes
optically thin.  The dip in their numbers near the midplane is then
explained because the optically thick dust layer obscures some
fraction of the stars along the line of sight.

To test this explanation, we built a simple
'toy model' galaxy with MS, RGB and AGB 
populations distributed as sech$^2$ profiles with the $z_{0}$ as shown
in Table~4.  All components were given identical radial
distributions with the $K_s$ band exponential scale length.
The dust component was assumed to also follow a sech$^2$ profile with a
variable scale height, $z_{\rm 0,dust}$, and a radial distribution
identical to that of the stars.  For simplicity we assumed that the dust 
has no effect at an optical depth less than one, but is completely
obscuring at greater optical depths.  Thus, along a line of sight, the
dust is completely transparent to $\tau = 1$, and completely opaque
beyond.  For each
vertical position we integrated the dust component along the line-of-sight
until an optical depth of one was reached, which set the depth of the
dust screen at that height.  Stellar density profiles for the
three separate populations were then created by totaling the number of
stars in front of the dust screen at each height.  We then normalized
the stellar density profiles as in Figure~5.

Figure~9 shows the resulting model of NGC~4144 for three values of
$z_{\rm 0,dust}$ presented for comparison to the observations shown in
Figure~5.  In each case the amount of dust in the midplane is the
same.  The underlying values for $z_0$ were adopted from Table~4 (374,
699, and 934 pc for the MS, AGB and RGB respectively).  The left panel
shows the results for a dust layer whose scale height is narrower than
all three stellar components.  For this case, there is a pronounced
dip in the surface density profile of all three components.  In the
middle panel, the value of $z_{\rm 0,dust}$ is between the $z_0$
values for the MS and the AGB/RGB populations.  In this case there is
a dip only in the AGB/RGB, because the height of the MS layer is
entirely confined within the opaque dust layer, allowing only the
unobscured stars on the near side of the galaxy to be detected.
The right panel has a dust layer larger than the both the MS and AGB
value and therefore a dip is seen only in the RGB component.  The
middle panel does a good job of qualitatively matching the
observations for NGC~4144 in Figure~5.

\begin{figure*}
\plotone{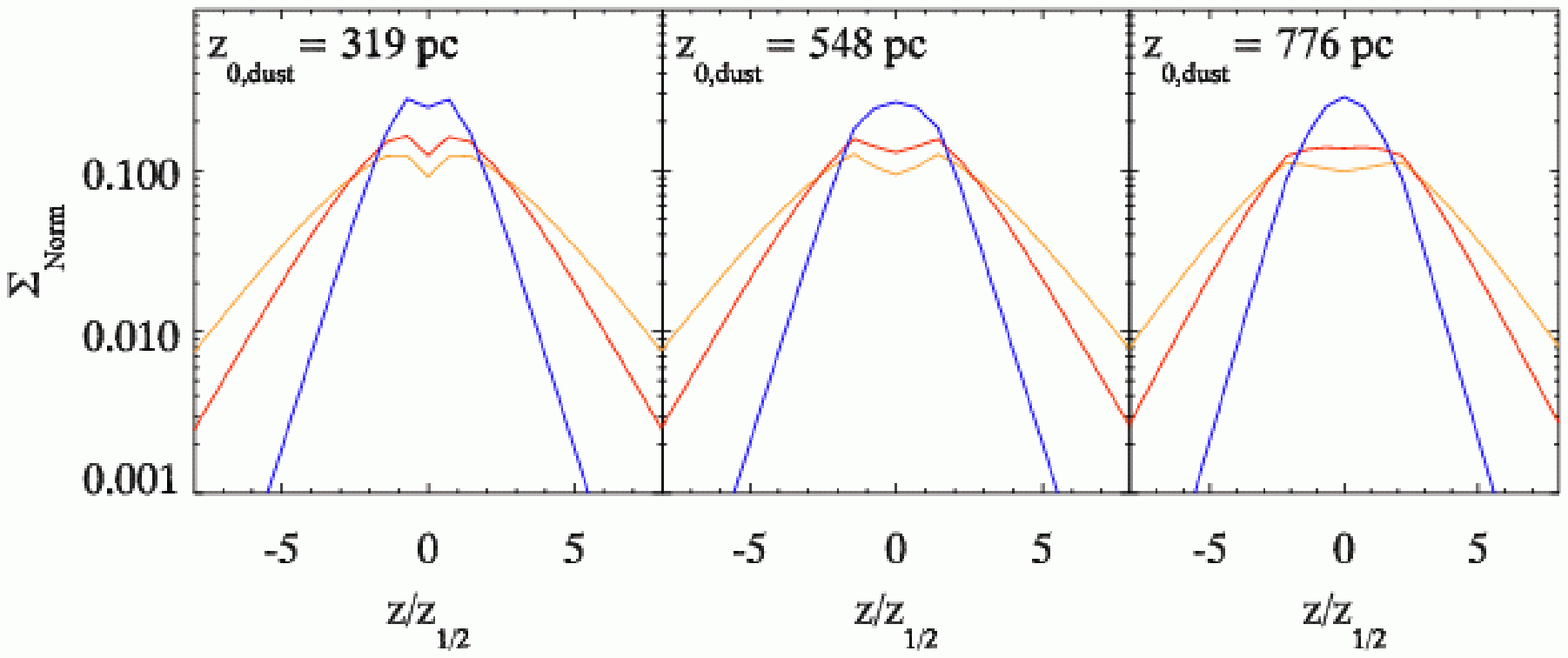}
\caption{Toy models including the effects of dust on the stellar
  density profiles of NGC~4144.  The three panels show different
  values of the dust scale height $z_{\rm 0,dust}$.  The blue, red,
  and orange/yellow lines in each panel represent the MS, AGB, and RGB
  profiles respectively.  Comparison to Figure~5 shows that the middle
  panel best matches the observations.}
\end{figure*}

Referring back to Figure~5, we can see that the main sequence
profile lacks a dip near the midplane for most of the
fields, while the RGB has a dip in all cases within 2-3~$z_{1/2}$.
This suggests that the dust in the galaxy is opaque below
2-3~$z_{1/2}$ and that it is distributed in a layer with thickness
greater than or equal to the MS stars.  Because the MS stars are
already distributed in a thicker distribution than in the Milky Way,
this result supports the \citet{dalcanton04} finding that galaxies
with circular velocities below 120 km~s$^{-1}$ have large dust scale
heights and do not form thin dust lanes.  All of the galaxies presented
here except NGC~4631 are below this circular velocity limit (Table~1).  

This model also suggests that although the depth to which we see in
each galaxy is different at differing scale heights, it is the same
for all the stellar populations at a single scale height.  This
validates the comparisons made in Fig.~6 between the ages of the
stellar populations at different scale heights.

Although this model matches the gross characteristics of many of the
profiles shown in Fig.~5, it fails to fully explain their details.
Most notably, in NGC~55 and NGC~4244, the MS profiles are
significantly lower than the best-fitting sech$^2$ function,
contradicting the idea that all the stars we are seeing lie in front
of a screen.  This is likely the result of our model's lack of
sophistication.  Physically, it doesn't take into account the
possibility of flares or changing scale lengths as a function of
stellar population.  In addition, it treats dust extinction in a very
simplistic fashion.  A more sophisticated treatment of the dust
\citep[such as the one presented by][]{matthews01} is beyond the scope
of this paper.  The conclusions reached in this
section should be considered tentative and will be tested in a later
paper, in which we will present a more realistic dust model.

%% One obvious problem with our model is that dust will not just remove
%% stars completely from our sample, but instead move them to the
%% lower-right - since we have defined regions to isolate different areas
%% of the CMD, the stars will be detectable until there is enough dust to
%% move them out of the defined CMD region.  This may be different for
%% our three regions - for instance, the presence of the offset peak in
%% NGC~4244 for the MS stars while the AGB stars peak near the midplane
%% might suggest that the dust removes more MS stars than AGB stars from
%% our CMD regions.  

\section{Metallicity Distribution Functions} \label{metallicitysec}

As has been shown previously
\citep[e.g.][]{dacosta90,armandroff93,frayn02}, the color of the red
giant branch near its tip can be used to constrain the metallicity of
old stellar populations.  Although reddening due to dust will prevent
an accurate measurement of the metallicity of stars near the midplane,
we can determine a rough metallicity for stars above the midplane
where the effect of dust and the contamination from AGB stars are
small (as shown in Figure~6, middle panel).

Figure~10 shows a composite CMD of all high-latitude stars (above
4$z_{1/2}$) in IC~5052, NGC~55, NGC~55-DISK, NGC~4144, NGC~4244, and
NGC~5023.  This disk height limit was chosen (1) to be well above any
of the dips associated with dust in Figure~5, (2) to dramatically
reduce contamination by AGB and HeB stars that might interfere in the
metallicity determination, and (3) to be where thick disk stars
dominate in the Milky Way \citep{chen01}.  Overlayed on the CMD are 10
Gyr old RGB isochrones \citep{girardi00} at metallicities ranging from
[Fe/H] $=-2.3$ to 0.0, with higher metallicities being redder.
Examination of Figure~10 shows that the peak of the distribution
falls between the [Fe/H] $=-1.3$ and -0.7 lines.  Roughly 13\% of
the stars fall bluewards of the [Fe/H] $=-2.3$ isochrone, probably due
to a combination of photometric error and the presence of AGB stars. 
Very few are redder than the [Fe/H] $=-0.4$ isochrone.  Overall,
Figure~10 indicates that most of the stars above 4$z_{1/2}$ are
moderately metal-poor.

\begin{figure}
\plotone{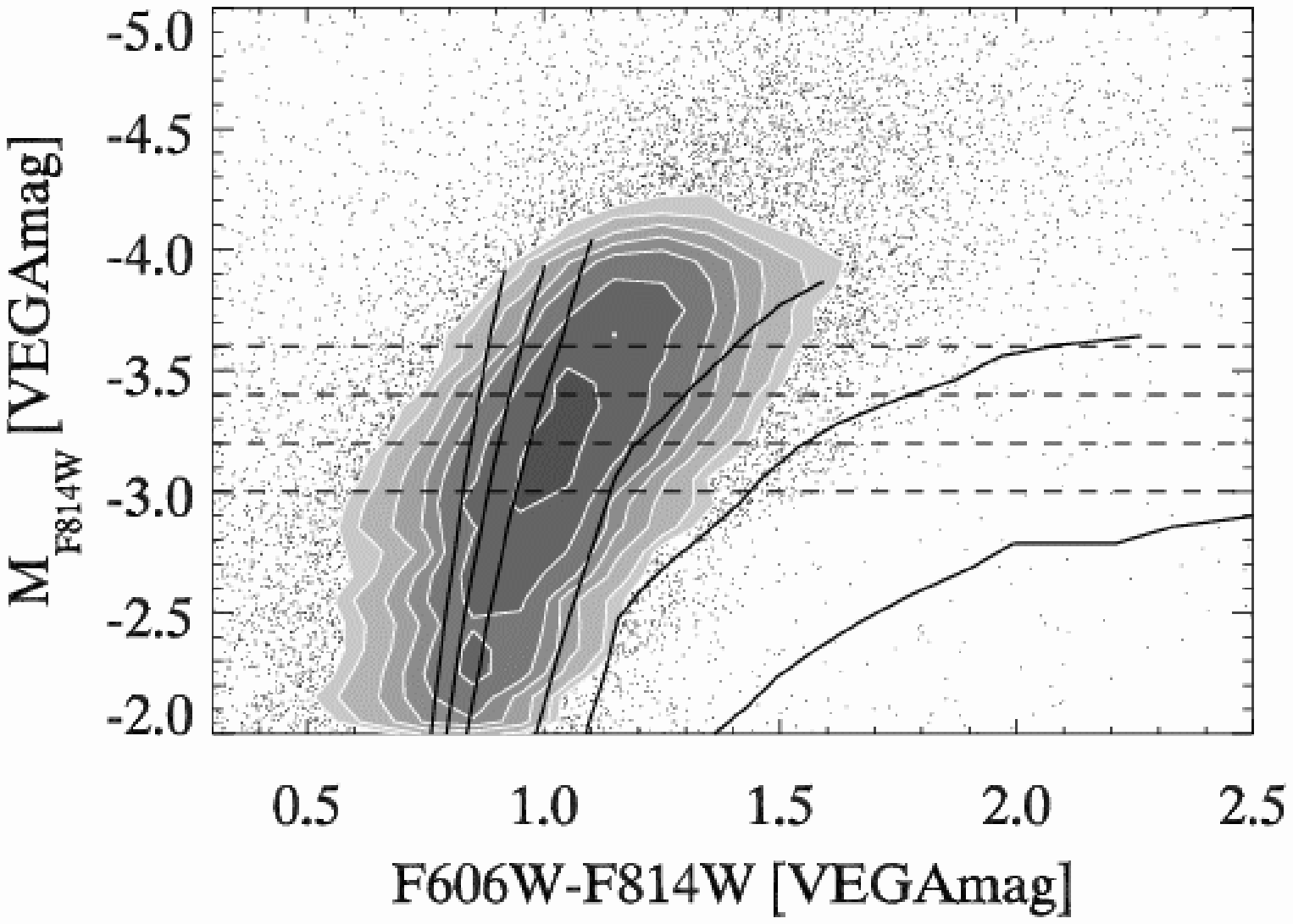}
\caption{Composite CMD showing the red giant branches of IC~5052,
  NGC~55, NGC~55-DISK, NGC~4144, NGC~4244, and NGC~5023 at disk
  heights $>$4$z_{1/2}$.  The absolute magnitude M$_{\rm F814W}$ was
  determined using the TRGB distance modulus in Table~1.  Solid lines
  show 10 Gyr old Padova models for the RGB 
  with [Fe/H] (from left to right) of -2.3, -1.7, -1.3, -0.7, -0.4,
  and 0.0.  The peak of the stellar distributions fall between
  [Fe/H] of -1.3 and -0.7.  Dashed lines indicate the three bins used
  to determine the metallicity distribution functions shown in Fig.~11.}
\end{figure}

Improving on the metallicity determinations in Paper~I, here we will
derive metallicity distribution functions (MDFs) using untransformed
magnitudes and Padova isochrones \citep{girardi00}.  To determine
metallicity distribution functions for individual galaxies we binned
the stars in up to three independent 0.2 magnitude wide bins centered
on M$_{\rm F814W}$ of -3.5, -3.3, and -3.1.  These are shown with
dashed lines in Figure~10.  The brightest bin was chosen to include
stars as metal-rich as [Fe/H] $=-0.4$, which do not get brighter than
M$_{\rm F814W}$ of -3.7.  We considered only the magnitude bins that
were above the 20\% completeness cutoff for crowded regions (thus
excluding the fainter bins in NGC~4144, see Figure~1).  This cut
eliminated NGC~4631 and NGC~4631-DISK from the analysis, because none
of the bins fell above their 20\% completeness limits.  We corrected
the colors and magnitudes of each star for foreground reddening, but
made no correction for completeness, since the change in completeness
across the color-range in question is similar to the error that would
be introduced by that correction.  We then determined the metallicity
of each star by linearly interpolating between the 10 Gyr isochrones
in each bin.  Since some of the stars bluewards of the [Fe/H] $=-2.3$
isochrone may be RGB stars scattered to bluer colors by photometric
error, we attempted to include these stars in the MDFs.  In each bin,
stars bluer than the [Fe/H] $=-2.3$ isochrone were given positive
color shifts by multiplying a Gaussian random number by their error.
This correction moved $\gtrsim$50\% of these stars redwards of the
[Fe/H] $=-2.3$ isochrone, and resulted in a slight increase
($\sim$0.01 in the normalized units) in the MDF between a [Fe/H] of
-1.3 and -2.3.  Table~5 gives the resulting number of stars used in
MDF determination, the median F606W-F814W error, and the peak and mean
metallicities of the MDFs.

\begin{deluxetable*}{lccccc}[h]
\tablewidth{0pt}
\tablecaption{Metallicity Determinations}
\tablehead{
     \colhead{Field}  &
     \colhead{Paper I} &
     \colhead{\# of stars} &
     \colhead{$\sigma$} &
     \colhead{MDF Peak} &
     \colhead{Mean} \\
     \colhead{}  &
     \colhead{[Fe/H]} &
     \colhead{} &
     \colhead{(F606W-F814W)} &
     \colhead{[Fe/H]} &
     \colhead{[Fe/H]} 
}
\startdata
IC 5052         &  -1.22 & 3770 & 0.21 & -0.7 & -0.9 \\ 
NGC 55          &  -1.41 & 1068 & 0.09 & -1.0 & -1.0 \\ 
NGC 55-DISK     &  -1.62 &  983 & 0.06 & -1.1 & -1.2 \\ 
NGC 4144        &  -1.50 & 1773 & 0.19 & -0.9 & -1.0 \\ 
NGC 4244        &  -1.45 & 1970 & 0.11 & -0.9 & -1.0 \\ 
NGC 4631        &  -1.25 &      &      &      &      \\ 
NGC 4631-DISK   &  -1.54 &      &      &      &      \\ 
NGC 5023        &  -1.71 & 1119 & 0.22 & -0.9 & -1.0 \\ 
\enddata
\tablecomments{The columns from left to right are: Field name, the
metallicity estimated in Paper I, the number of stars used in the
metallicity determination in this paper, the typical color error of
those stars, the peak of the metallicity distribution function (from
Fig.~11), and the mean metallicity.}
\end{deluxetable*}

Figure~11 shows the resulting MDFs for each field.  The shaded regions
show the total range in the MDFs as derived in the different magnitude
bins, and indicate that our results are consistent among the different
magnitude ranges.  In general, the MDFs peak at metallicities [Fe/H]
of -0.7 to -1.1 as expected from Figure~10, and have a tail of stars
to low metallicities.

Before analyzing the MDFs further, we note that there are several
uncertainties in the detailed shapes of the metallicity
distribution functions.  First, 
they are based on isochrone models and not empirical data \citep[as
in][]{sarajedini01}.  Second, the color errors on the metal-poor end
(Table~5) translate into a large error in metallicity.  We estimated
the errors as a function of metallicity by inserting stars of a
specific metallicity/color, giving them appropriate color errors, and
then determining the spread in the resulting metallicity distribution.
This procedure gave errors of 0.5-0.8 dex at [Fe/H] $=-2.3$.  However,
the shape of the distributions are much more believable on the
metal-rich end where the isochrones are well separated, giving errors
of less than 0.1 dex at [Fe/H] $=-0.4$.  At the peak of the MDF
([Fe/H] $\sim$ -0.9), typical errors are 0.2 dex, suggesting that the
peak metallicities derived here are fairly reliable.

Comparing the peak metallicity of the extraplanar stars to known Milky Way
populations, Figure~11 indicates that the metallicities of the
extraplanar stars are a factor of ten times too high to be analogs of
the Milky Way's stellar halo \citep[${\rm [Fe/H]} \sim-1.7$,][]{wyse95}.
This result would not change even if all the stars bluewards of the
[Fe/H] $=-2.3$ isochrone are low metallicity RGB stars.  We note
however, that the low metallicity of the Milky Way halo may not be
typical.  The halo of Andromeda has been found to be much more
metal-rich than in the Milky Way, with a peak [Fe/H] $\sim$ -0.6
\citep{holland96,brown03}, although there is difficulty ascribing
these outer stars to a halo population {\it{per se}}, given M31's
complicated outer structure \citep{ferguson02}.
%while
%observations in the outer (thick?)  disk show a population of [Fe/H]
%$\sim$ -0.1 red giants. 

Of all the Milky Way components, we find that the peak metallicities
are most consistent with those of the metallicity of the Milky Way
thick disk, which has [Fe/H] $\sim$ -0.8 based on F/G dwarfs
\citep{wyse95}.  The extraplanar stars studied here are somewhat more
metal poor than the Milky Way's thick disk (by up to 0.3 dex).  However,
this offset may not be surprising given the lower mass of our galaxy
sample ($V_{c}\sim80$ km~s$^{-1}$ vs $V_{c}\sim220$ km~s$^{-1}$ for
the Milky Way). 

As in the Milky Way \citep{wyse95,haywood01}, the metallicities of the
extraplanar stars appear to be more metal poor than the thin, young,
main sequence population.  Although dust prevents us from reliably
measuring the metallicity of stars near the midplane, we can estimate
their metallicity using the current gas phase metallicity.  NGC 55,
the only galaxy in our sample with a gas phase abundance measurement,
has $12+\log{\rm (O/H)}=8.32$ at one disk scale length \citep{garnett02}.
This metallicity corresponds to [Fe/H] $\sim -0.6$ (assuming [Fe/O]
$=0$), which is 0.5 dex more metal rich than the extraplanar stars at
a comparable radius.  Other late-type disks with similar rotational
velocities from the \citet{garnett02} compilation have comparable
gas-phase metallicities, suggesting that the offset in metallicity
between the midplane and the extraplanar populations is likely to be
systematic.

Although they do not explicitly examine stars as a function of scale
height, studies of metallicities in other galaxies using methods
similar to ours also find broad agreement with the presence of an
extended [Fe/H] $\sim-1$ population of RGB stars.  The LMC (which has a
mass similar to the galaxies in our sample) has a peak metallicity
distribution of [Fe/H] $\sim-0.6$ for RGB stars in the disk
\citep{cole00}.  Recent papers on the outer regions of M33 also find
peak [Fe/H] values of -1.0 \citep{davidge03,tiede04}.  Furthermore, a
recent paper by \citet{davidge05}, derives a [Fe/H] of roughly -1 for
NGC~55 using near-IR photometry of resolved extraplanar stars, closely
matching our peak metallicity in Figure~11.  Our data and others
therefore suggest the pervasive presence of a significant
[Fe/H] $\sim-1$ old population in late-type galaxies.  If our
association of this population with a thicker disk is generally true
in other galaxies, then it presents an attractive solution to the
``G-dwarf'' problem seen in the Milky Way by providing the necessary
prompt initial enrichment for stars in the thin disk \citep{truran71}.

Overplotted on Figure~11 as dashed lines are the expected metallicity
distributions for closed-box ``simple'' chemical evolution models
\citep[Eq. 20 of][]{pagel97} scaled to the peaks of the MDFs.  While
the basic shape of these models are similar to our MDFs, there appears
to be a deficit in some galaxies of stars at both low and high
metallicities.  A deficit at high metallicities is expected if
star formation truncates before all the gas is consumed.  Within the
context of thick disk formation models, this truncation may occur if
some of the gas reservoir that forms the extraplanar stars instead
settles into the thin disk, if the extraplanar stars were heated from
a previously thin but gas-rich disk, or if the extraplanar stars were
directly accreted from merging satellites that suffered from tidal
stripping or supernova blowout.  The apparent deficit of stars at low
metallicities may be another manifestation of the widespread G-dwarf
problem.  Thus, while the existence of a substantial population of
stars at [Fe/H] $\sim-1$ may help to solve the G-dwarf problem in the
thin disk, it may have simply pushed the problem into a new component.
The solution to the extraplanar G-dwarf problem will
likely lie among the suite of popular models previously explored for
the thin disk \citep[see][]{pagel97}.  However, some of the deficit of stars
at low metallicities may also result from the photometric errors and
methods used to construct the MDFs, as discussed above.

%% From Skillman et al 1989:  assume [Fe/O]=0, and O/H_solar=8.3e-4 
%% (Lambert 1978), which gives 12+log(O/H_solar)=8.919.  Garnett 2002
%% gives 12+log(O/H)=8.32 for NGC 55, giving [Fe/H]=-0.6, and for two
%% other spirals with V_c=90, 8.56 and 8.26 at R_eff ([Fe/H]=-0.36, -0.66)

Finally, we note that the peaks of the metallicity distribution
functions given in Table~5 are significantly more metal-rich than our
previous determination presented in Paper~I.  Rather than using native
F606W$-$F814W colors, these earlier measurements applied the
metallicity-color relation of \citet{lee93} to the mean color of the
giant branch transformed to the Johnson-Cousins filter system.  These
previous values, reproduced in Table~5, range from [Fe/H] of -1.2 to
-1.7, versus -0.7 to -1.1 in the present work.  We believe that in
addition to the magnitude transformation, the difference in derived
metallicity results in part from the  difference in binning (in [Fe/H]
vs.\ F606W-F814W), such that the mean color does not correspond to the
peak metallicity.  This offset can be seen in the 0.1-0.2 dex offset
between the mean and peak metallicity in Table~5.  Furthermore, the
Paper~I determination also 
included stars at somewhat lower disk heights, thereby increasing the
number of AGB contaminants.  We believe that the MDFs and their peak
metallicity that we present here are more reliable than the estimate
given in Paper~I.  

%% The metallicity we have determined for the RGB stars at high latitude
%% appears to match well to the metallicites of the F and G stars in the
%% Milky Way thick disk and is significantly poorer than thin disk F and
%% G stars \citep{wyse95}.  This might seem to contradict our claim that
%% these stars make up an old thin disk.  However, \citet{garnett02}
%% shows that the current day metallicity of galaxies at velocities
%% $\sim$100 km/sec is 0.4 to 0.6 dex more metal-poor than the Milky
%% Way.  The metallicities of the RGB stars are therefore only slightly
%% more metal poor than the current stellar population.  In addition,
%% many of the long-lived stars used to locally determine the Milky Way
%% thin disk metallicity may be young (and therefore somewhat more
%% metal-rich) as suggested by \citep{haywood01}.

\begin{figure*}
\plotone{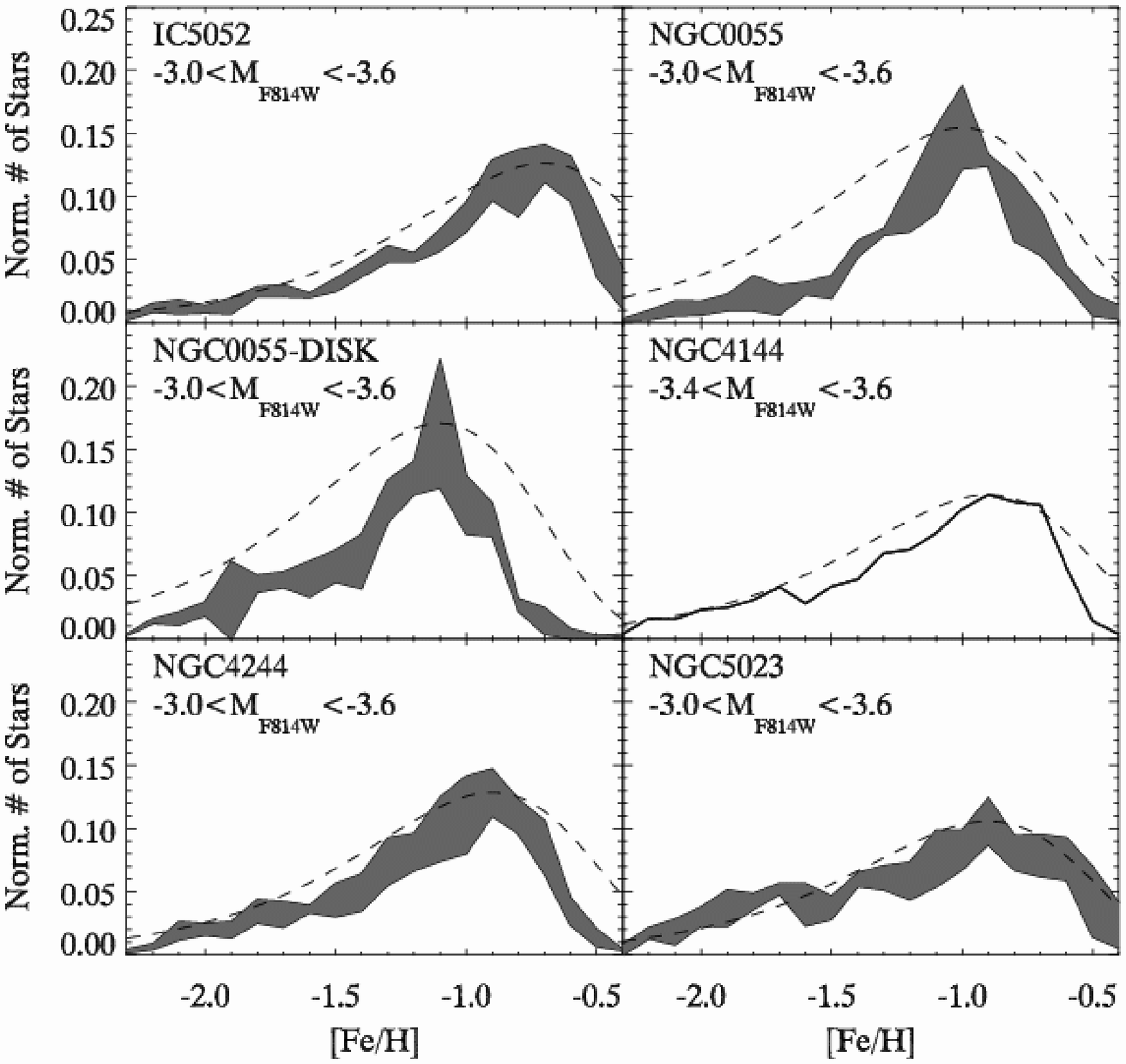}
\caption{Metallicity distribution functions for stars at disk heights
  $>$4$z_{1/2}$.  Gray regions indicate the range of values obtained in
  different absolute magnitude bins.  The bins used for this
  determination are 0.2 magnitudes wide and are shown in Fig.~10.  Only
  bins above the magnitude limits in Table~3 were used.  Dashed lines
  indicate closed-box chemical evolution models scaled to the peak of
  the MDF.}
\end{figure*}

\subsection{Vertical Metallicity \& Color Gradients} \label{vertgrad}

Models for the origin of extraplanar stars (i.e.\ disk heating, direct
accretion, etc.) predict different degrees of variation in the stellar
metallicity with height above the plane.  To investigate the vertical
variation of metallicity with disk height $z$ we have examined the
median color of the RGB stars as a function of the height above the
midplane.  Figure~12 shows the median color of RGB stars between
M$_{\rm F814W}$ of -3.2 and -3.6 and redwards of F606W-F814W=0.7,
binned by the scale height of the galaxies.  Data are plotted where
errors in the median color are $<$0.05 magnitudes.  The hatched region
at low disk heights shows where the effects of internal reddening may
impact the colors of the stars.  For the three fields with profiles
extending beyond 2-3~kpc we bin the RGB stars at large disk heights in
a single bin to reach adequate signal-to-noise in our measurement of
the median color, plotted as diamonds in Figure~12.  We note that
these are the same stars that comprise the possible extended components
discussed in \S4.2.

Figure~12 demonstrates three main points.  First, the color gradients
in the galaxies are relatively small, indicating that the stars have
nearly uniform metallicity with increasing distance above the plane,
particularly at scale heights above the region potentially affected by
dust.  However, we note that the stars at very large radii (shown with
diamonds) do tend towards bluer colors, possibly indicating the
presence of a more metal-poor population at $z \gtrsim 10 z_{1/2}$
(2-3 kpc).  Second, the color gradients show no systematic trends, and
are equally likely to be rising or falling.  Finally, all the galaxies
have very similar RGB colors (as demonstrated already in Figure~11).
%% We note that the offset in the color of IC~5052 (our lowest latitude
%% galaxy) from the rest of the galaxies could be explained if the
%% foreground reddening were $\sim$2-3 times the \citet{schlegel98} value
%% of E(B-V)$=0.051$ mag.  However, this is well above the expected error
%% in the reddening.

Our metallicity gradient results are consistent with previous
observations of these and other low-mass spiral galaxies
\citep{davidge05,tikhonov05a,mould05}.  \citet{mould05} used HST
archival data to study the vertical properties of disks in four
low-luminosity ($M_V \sim -16$) edge-on galaxies comparable to those
studied here.  Using AGB, RGB, and red supergiant stars over a large
range of magnitudes ($-8.5\lesssim M_I\lesssim-1.5$), he calculates
the mean colors up to 2 kpc from the plane.  His main results are that
there are slight or no color gradients as a function of disk height
and that the metallicities of the stars at disk heights between 400
and 2000 pc are between -0.8 and -1.0 in all four galaxies, in
excellent agreement with what we find here.  \citet{tikhonov05a} also
state that any metallicity gradient in NGC~55 is very small, in
agreement with our observed lack of color-gradients in Figure~12.
Recent simulations of thick disks by \citet{brook05} also show a lack
of any metallicity gradient with disk height.

The lack of strong metallicity gradients can be explained in a number
of ways.  \citet{mould05} suggests the lack of metallicity gradients
rules out dissipative and simple accretion models for thick disk
formation, and favors a model in which thick disks form during
interactions.  We note that the lack of metallicity gradient may also
be enhanced by an ``age-bias'' in the metallicity measurements.  The
mass of stars on the RGB changes from $\sim$2 M$_\odot$ at an age of 1
Gyr to $<$1 M$_\odot$ at 10 Gyr.  Given the steepness of the IMF in
this mass range, the RGB age distribution will be weighted towards
older ages, and therefore to a more uniform metallicity.  However,
based on the constant star formation rate models in \S4.1, it still
appears that the RGB stars could span a wide range of ages, even in
the presence of the expected age-bias.  Therefore, the simplest
interpretation of the lack of gradients in our data is that many of
the RGB stars at disk heights above 4$z_{1/2}$ ($\sim$1 kpc) formed at
a similar time and thus have comparable enrichment histories,
eliminating any metallicity gradient.  This scenario could be
explained either by a sudden heating of the disk by interaction, or by
accretion of gas-rich satellites which resulted in the formation of a
thick component.  N-body simulations have shown that early merging and
accretion events can produce thick disks with old ages
\citep{abadi03,brook04,brook05}.

\begin{figure}
\plotone{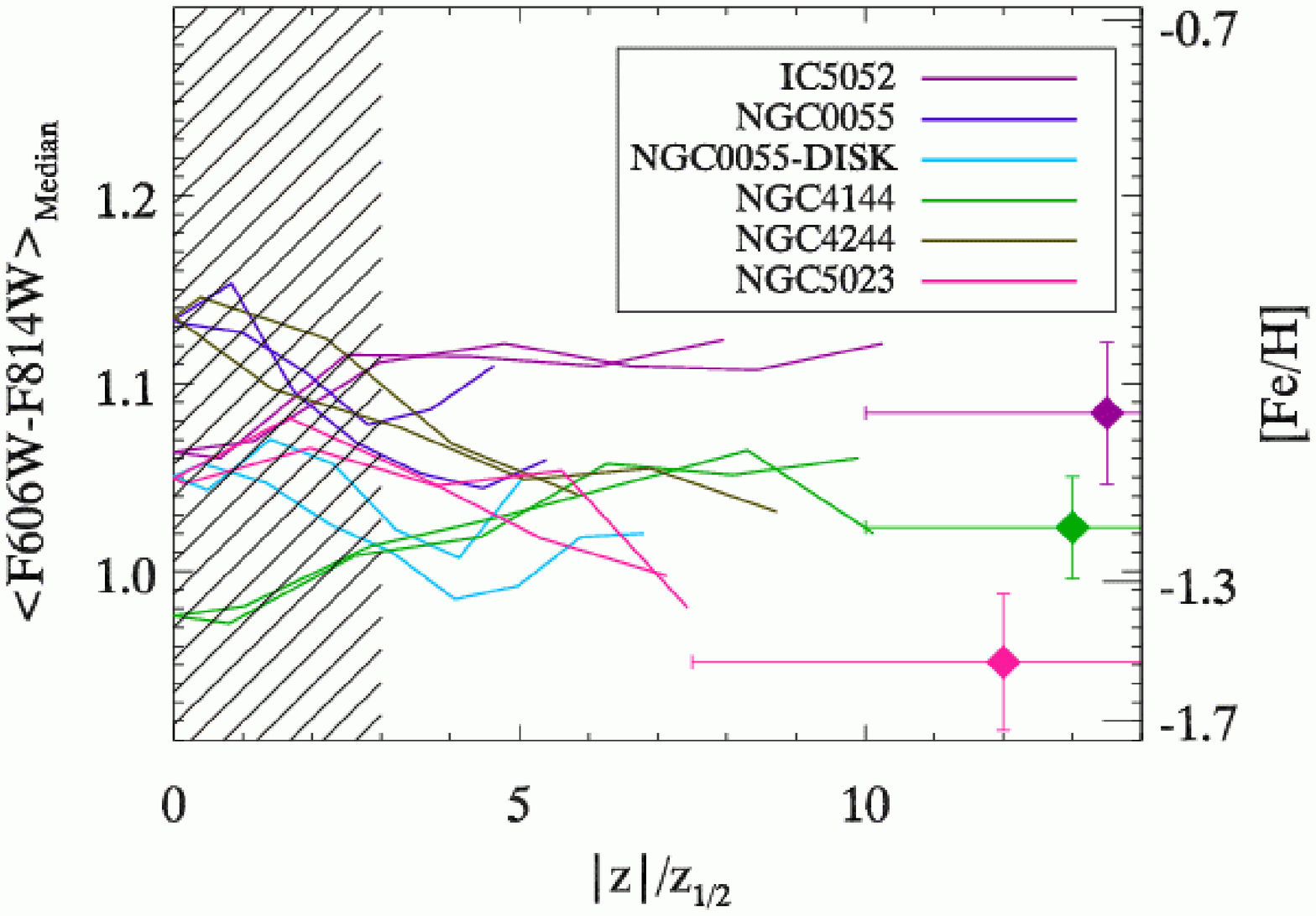}
\caption{The median RGB color as a function of disk height in each
  galaxy.  This includes stars with F606W-F814W redder than 0.7 and
  M$_{\rm F814W}$ between -3.2 and -3.6. The error on all plotted data
  is $<$0.05 magnitudes.  The diamonds indicate the median color of
  all stars at large disk heights and is plotted only for the three
  galaxies with significant numbers of stars at large scale heights.
  The hatched area indicates the heights at which internal reddening
  may affect the stellar colors.  The right axis shows the mean RGB
  color of Padua isochrones \citep{girardi04} at three different
  metallicities.}
\end{figure}

\section{Discussion \& Conclusions}    \label{discussionsec}

The work presented here has identified a number of main observational
results:

\begin{itemize}

\item In low mass, late-type galaxies the thickness of a stellar
  population increases systematically with the age of the stars being
  studied.  This behavior has been seen not just in all six of the
  galaxies studied here, but in all other HST studies of edge-on
  late-type galaxies, and in the Milky Way as well.  The larger scale
  heights of older stellar populations is therefore likely to be a
  generic property of galaxy disks.

\item All of the studied galaxies show a clear intermediate age
  ($1-5$~Gyr old) population whose scale height is intermediate
  between that of the young main sequence stars and the older red
  giants.

\item The metallicity of the dominant old stellar population has
  [Fe/H] $\sim -1$, but shows little or no gradient between $3z_{1/2}$
  and $10z_{1/2}$ above the plane.  Above this height ($>\!2-3$~kpc),
  there are tentative indications of decreasing metallicities, which
  may be associated with slight overdensities in the RGB surface
  density at similar distances above the midplane.

\item In the low mass galaxies studied here ($V_c\!\sim\!70-130$~km~sec$^{-1}$),
  the young stellar population is systematically thicker than in the
  MW, and has a vertical scale height comparable to the thickness of
  the dust layer.  This suggests that the cold ISM has a larger scale
  height in low mass galaxies, consistent with the lack of dust lanes
  observed in such systems \citep{dalcanton04}.

\item The young and intermediate-age stellar populations dominate the
  integrated {\emph{near-infrared}} light of late-type low mass
  galaxies.

\end{itemize}

We now interpret these observational facts in the context of disk
formation models.  First, taken at face value, the old RGB component's
$\sim$3:1 axial ratio and [Fe/H] $\sim -1$ metallicity suggest a
close correspondence with the Milky Way's thick disk.  However, each
of our galaxies' scale heights steadily increase from the young main
sequence to the intermediate age AGB and the older RGB.  The
uniformity of this trend strongly suggests that our disks are not
simply the superposition of two components (i.e.\ a thick and thin
disk).  Instead, the data require a more complex model incorporating
some disk heating to explain the systematically larger scale height of
the intermediate-age population.  The necessary disk heating would
also have affected any older population, and thus must make some
contribution to the thicker population of RGB stars.  The required
amount of disk heating is much smaller than is seen in the Milky Way
(\S4.3), and could likely be provided by molecular clouds or minor
mergers.  The latter scenario is slightly favored by the large
variations in the apparent change of disk scale height with time
(Figures~6~\&~8).  Heating through satellite accretion and interaction
could naturally produce these stochastic variations.  However,
numerical simulations of heating in diffuse low mass disks would be
required to definitively constrain any of the above scenarios and to
assess how significant a contribution heating may have made to the thicker
RGB component.

While the above argument strongly suggests that disk heating must play
some role in the production of the extraplanar stars, the lack of a
metallicity gradient in RGB stars at moderate disk heights
($500-2000$~pc, as shown here and in Mould 2005) suggests that
steady disk heating cannot entirely explain the thickest component of
old RGB stars.  If the past star formation rate has been constant,
then there is a significant overlap between the stellar ages of the
AGB and RGB regions of the color-magnitude diagram (Figure~4). A
significant fraction of the RGB stars should therefore have smaller
scale heights, younger ages, and enriched metallicities, and would thus
produce a steady increase in metallicity towards the midplane for all
but the most contrived scenarios.  The most attractive explanation for
the lack of metallicity gradient is that instead of a constant star
formation rate, the majority of RGB stars at all scale heights must
have formed early and with a well-mixed metallicity distribution.
Such a population would dwarf any subsequent population of enriched
RGB stars with lower scale heights.  While steady dynamical heating
could push this ancient population to larger scale heights, it could
not simultaneously account for recent dynamical observations of
counter-rotating disks at these disk heights in comparable galaxies
\citep{yoachim05}.  Taken together, these observations are better
explained by scenarios involving the formation of a thick disk of
stars in merger events \citep[as in][]{abadi03,brook04,brook05}.  Overall, our
results require that some disk heating occurs at intermediate ages (to
puff up the AGB component), but that events at earlier times
(interactions or mergers) created a majority of the RGB stars over a
short timescale.

%{\bf{(XXX I feel like this paragraph repeats the revised text in
%    4.2.1, and the bullet points.  To make it distinct, it might be
%    better reworked into a more general ``evidence for stellar halos''
%    discussion.)}}  
Finally, we present tenuous evidence for an extended old component
seen only at disk heights $>$ 2-3 kpc.  At large scale heights we see
marginal overdensities of stars in the RGB profiles of Fig.~7.  There
also seems to be a reduction in the metallicity of RGB stars at this
height (Fig.~12).  In one of our galaxies, \citet{tikhonov05a} finds
strong evidence for a very extended component of RGB stars extending
from $\sim$2-7~kpc.  While this component appears to have an
exponential distribution, its z$_0$ value of $\sim$2 kpc (compared to
a radial scale length of $\sim$1 kpc) strongly suggests it is not a
disk.  These extended components are detected at about the same height
where the halo becomes prominent in the stacked Sloan images of
\citet{zibetti04a}.  However, based on our observations, we are unable
to assess the properties or frequency of these components.

The present-day structure of galaxy disks results from a complex
mixing of effects and a full explanation requires detailed knowledge
of the the star formation history, merging events, and disk heating.
Studies like the one presented here help to disentangle these effects
and determine their relative importance as function of galaxy type and
mass.  This study also shows the promise that HST observations of
resolved stars have for enabling the detailed analysis of low surface
brightness stellar components in galaxies outside the Local Group.  A
comparison of our data with N-body simulations of low mass disk
galaxies would assist in constraining disk heating and merging
scenarios.  Unfortunately, current simulations of disk galaxy
formation have focused on massive galaxies like the Milky Way
\citep{abadi03,brook04,brook05}.  Also, deeper observations that fully
resolve the red and blue horizontal branches would greatly improve
constraints on the star formation histories of these galaxies and
improve our understanding of their structure.

Acknowledgements: The authors would like to thank Andrew Dolphin and
Antonio Aparicio for their help in generating synthetic CMDs, Leo
Girardi for supplying us with isochrones, our anonymous referee for
their thoughtful suggestions, and Peter Yoachim, Andrew West, and
Kevin Covey for helpful discussions.  This work was supported by
HST-GO-09765, the Sloan foundation, and NSF Grant CAREER AST-0238683.


\begin{thebibliography}{98}
\expandafter\ifx\csname natexlab\endcsname\relax\def\natexlab#1{#1}\fi

\bibitem[{{Abadi} {et~al.}(2003){Abadi}, {Navarro}, {Steinmetz}, \&
  {Eke}}]{abadi03}
{Abadi}, M.~G., {Navarro}, J.~F., {Steinmetz}, M., \& {Eke}, V.~R. 2003, \apj,
  597, 21

\bibitem[{{Aoki} {et~al.}(1991){Aoki}, {Hiromoto}, {Takami}, \&
  {Okamura}}]{aoki91}
{Aoki}, T.~E., {Hiromoto}, N., {Takami}, H., \& {Okamura}, S. 1991, \pasj, 43,
  755

\bibitem[{{Aparicio} \& {Gallart}(2004)}]{aparicio04}
{Aparicio}, A., \& {Gallart}, C. 2004, \aj, 128, 1465

\bibitem[{{Armandroff} {et~al.}(1993){Armandroff}, {Da Costa}, {Caldwell}, \&
  {Seitzer}}]{armandroff93}
{Armandroff}, T.~E., {Da Costa}, G.~S., {Caldwell}, N., \& {Seitzer}, P. 1993,
  \aj, 106, 986

\bibitem[{{Barbanis} \& {Woltjer}(1967)}]{barbanis67}
{Barbanis}, B., \& {Woltjer}, L. 1967, \apj, 150, 461

\bibitem[{{Becker} {et~al.}(1988){Becker}, {Mebold}, {Reif}, \& {van
  Woerden}}]{becker88}
{Becker}, R., {Mebold}, U., {Reif}, K., \& {van Woerden}, H. 1988, \aap, 203,
  21

\bibitem[{{Bekki} \& {Chiba}(2001)}]{bekki01}
{Bekki}, K., \& {Chiba}, M. 2001, \apj, 558, 666

\bibitem[{Bensby {et~al.}(2005)Bensby, Feltzing, Lundstrom, \&
  Ilyin}]{bensby05}
Bensby, T., Feltzing, S., Lundstrom, I., \& Ilyin, I. 2005, Astron. Astrophys.,
  433, 185

\bibitem[{{Benson} {et~al.}(2004){Benson}, {Lacey}, {Frenk}, {Baugh}, \&
  {Cole}}]{benson04}
{Benson}, A.~J., {Lacey}, C.~G., {Frenk}, C.~S., {Baugh}, C.~M., \& {Cole}, S.
  2004, \mnras, 351, 1215

\bibitem[{{Bertelli} {et~al.}(1994){Bertelli}, {Bressan}, {Chiosi}, {Fagotto},
  \& {Nasi}}]{bertelli94}
{Bertelli}, G., {Bressan}, A., {Chiosi}, C., {Fagotto}, F., \& {Nasi}, E. 1994,
  \aaps, 106, 275

\bibitem[{{Binney} {et~al.}(2000){Binney}, {Dehnen}, \& {Bertelli}}]{binney00}
{Binney}, J., {Dehnen}, W., \& {Bertelli}, G. 2000, \mnras, 318, 658

\bibitem[{Brook {et~al.}(2005)Brook, Gibson, Martel, \& Kawata}]{brook05}
Brook, C.~B., Gibson, B.~K., Martel, H., \& Kawata, D. 2005

\bibitem[{{Brook} {et~al.}(2004){Brook}, {Kawata}, {Gibson}, \&
  {Freeman}}]{brook04}
{Brook}, C.~B., {Kawata}, D., {Gibson}, B.~K., \& {Freeman}, K.~C. 2004, \apj,
  612, 894

\bibitem[{{Brown} {et~al.}(2003){Brown}, {Ferguson}, {Smith}, {Kimble},
  {Sweigart}, {Renzini}, {Rich}, \& {VandenBerg}}]{brown03}
{Brown}, T.~M., {Ferguson}, H.~C., {Smith}, E., {Kimble}, R.~A., {Sweigart},
  A.~V., {Renzini}, A., {Rich}, R.~M., \& {VandenBerg}, D.~A. 2003, \apjl, 592,
  L17

\bibitem[{{Buonanno} {et~al.}(1994){Buonanno}, {Corsi}, {Buzzoni}, {Cacciari},
  {Ferraro}, \& {Fusi Pecci}}]{buonanno94}
{Buonanno}, R., {Corsi}, C.~E., {Buzzoni}, A., {Cacciari}, C., {Ferraro},
  F.~R., \& {Fusi Pecci}, F. 1994, \aap, 290, 69

\bibitem[{{Burstein}(1979)}]{burstein79}
{Burstein}, D. 1979, \apj, 234, 829

\bibitem[{{Carlberg}(1987)}]{carlberg87}
{Carlberg}, R.~G. 1987, \apj, 322, 59

\bibitem[{{Carlberg} \& {Sellwood}(1985)}]{carlberg85}
{Carlberg}, R.~G., \& {Sellwood}, J.~A. 1985, \apj, 292, 79

\bibitem[{{Chen} {et~al.}(2001){Chen}, {Stoughton}, {Smith}, {Uomoto}, {Pier},
  {Yanny}, {Ivezi{\' c}}, {York}, {Anderson}, {Annis}, {Brinkmann}, {Csabai},
  {Fukugita}, {Hindsley}, {Lupton}, {Munn}, \& {the SDSS
  Collaboration}}]{chen01}
{Chen}, B., {Stoughton}, C., {Smith}, J.~A., {Uomoto}, A., {Pier}, J.~R.,
  {Yanny}, B., {Ivezi{\' c}}, {\v Z}., {York}, D.~G., {Anderson}, J.~E.,
  {Annis}, J., {Brinkmann}, J., {Csabai}, I., {Fukugita}, M., {Hindsley}, R.,
  {Lupton}, R., {Munn}, J.~A., \& {the SDSS Collaboration}. 2001, \apj, 553,
  184

\bibitem[{{Cole} {et~al.}(2000){Cole}, {Smecker-Hane}, \& {Gallagher}}]{cole00}
{Cole}, A.~A., {Smecker-Hane}, T.~A., \& {Gallagher}, J.~S. 2000, \aj, 120,
  1808

\bibitem[{{Da Costa} \& {Armandroff}(1990)}]{dacosta90}
{Da Costa}, G.~S., \& {Armandroff}, T.~E. 1990, \aj, 100, 162

\bibitem[{{Dalcanton} \& {Bernstein}(2002)}]{dalcanton02}
{Dalcanton}, J.~J., \& {Bernstein}, R.~A. 2002, \aj, 124, 1328

\bibitem[{{Dalcanton} {et~al.}(2004){Dalcanton}, {Yoachim}, \&
  {Bernstein}}]{dalcanton04}
{Dalcanton}, J.~J., {Yoachim}, P., \& {Bernstein}, R.~A. 2004, \apj, 608, 189

\bibitem[{Davidge(2005)}]{davidge05}
Davidge, T. 2005, astro-ph/0501173

\bibitem[{{Davidge}(2003)}]{davidge03}
{Davidge}, T.~J. 2003, \aj, 125, 3046

\bibitem[{{de Grijs} \& {Peletier}(1997)}]{degrijs97}
{de Grijs}, R., \& {Peletier}, R.~F. 1997, \aap, 320, L21

\bibitem[{{de Grijs} \& {van der Kruit}(1996)}]{degrijs96}
{de Grijs}, R., \& {van der Kruit}, P.~C. 1996, \aaps, 117, 19

\bibitem[{{De Simone} {et~al.}(2004){De Simone}, {Wu}, \&
  {Tremaine}}]{desimone04}
{De Simone}, R., {Wu}, X., \& {Tremaine}, S. 2004, \mnras, 350, 627

\bibitem[{{Dolphin}(2002)}]{dolphin02}
{Dolphin}, A.~E. 2002, \mnras, 332, 91

\bibitem[{{Eggen} {et~al.}(1962){Eggen}, {Lynden-Bell}, \& {Sandage}}]{eggen62}
{Eggen}, O.~J., {Lynden-Bell}, D., \& {Sandage}, A.~R. 1962, \apj, 136, 748

\bibitem[{{Feltzing} {et~al.}(2003){Feltzing}, {Bensby}, \& {Lundstr{\"
  o}m}}]{feltzing03}
{Feltzing}, S., {Bensby}, T., \& {Lundstr{\" o}m}, I. 2003, \aap, 397, L1

\bibitem[{{Ferguson} {et~al.}(2002){Ferguson}, {Irwin}, {Ibata}, {Lewis}, \&
  {Tanvir}}]{ferguson02}
{Ferguson}, A.~M.~N., {Irwin}, M.~J., {Ibata}, R.~A., {Lewis}, G.~F., \&
  {Tanvir}, N.~R. 2002, \aj, 124, 1452

\bibitem[{{Florido} {et~al.}(2001){Florido}, {Battaner}, {Guijarro}, {Garz{\'
  o}n}, \& {Jim{\' e}nez-Vicente}}]{florido01}
{Florido}, E., {Battaner}, E., {Guijarro}, A., {Garz{\' o}n}, F., \& {Jim{\'
  e}nez-Vicente}, J. 2001, \aap, 378, 82

\bibitem[{{Frayn} \& {Gilmore}(2002)}]{frayn02}
{Frayn}, C.~M., \& {Gilmore}, G.~F. 2002, \mnras, 337, 445

\bibitem[{{Freeman} \& {Bland-Hawthorn}(2002)}]{freeman02}
{Freeman}, K., \& {Bland-Hawthorn}, J. 2002, \araa, 40, 487

\bibitem[{{Freeman}(1991)}]{freeman91}
{Freeman}, K.~C. 1991, in Dynamics of Disc Galaxies, Proceedings of Varberg
  Conference, ed. B. Sundelius, 15--+

\bibitem[{{Fry} {et~al.}(1999){Fry}, {Morrison}, {Harding}, \&
  {Boroson}}]{fry99}
{Fry}, A.~M., {Morrison}, H.~L., {Harding}, P., \& {Boroson}, T.~A. 1999, \aj,
  118, 1209

\bibitem[{{Garnett}(2002)}]{garnett02}
{Garnett}, D.~R. 2002, \apj, 581, 1019

\bibitem[{{Gilmore}(1984)}]{gilmore84}
{Gilmore}, G. 1984, \mnras, 207, 223

\bibitem[{{Gilmore} \& {Reid}(1983)}]{gilmore83}
{Gilmore}, G., \& {Reid}, N. 1983, \mnras, 202, 1025

\bibitem[{{Gilmore} {et~al.}(2002){Gilmore}, {Wyse}, \& {Norris}}]{gilmore02}
{Gilmore}, G., {Wyse}, R.~F.~G., \& {Norris}, J.~E. 2002, \apjl, 574, L39

\bibitem[{{Girardi}(2004)}]{girardi04}
{Girardi}, L. 2004, private communication

\bibitem[{{Girardi} {et~al.}(2000){Girardi}, {Bressan}, {Bertelli}, \&
  {Chiosi}}]{girardi00}
{Girardi}, L., {Bressan}, A., {Bertelli}, G., \& {Chiosi}, C. 2000, \aaps, 141,
  371

\bibitem[{{Gnedin}(2003)}]{gnedin03}
{Gnedin}, O.~Y. 2003, \apj, 589, 752

\bibitem[{{Gratton} {et~al.}(2003){Gratton}, {Carretta}, {Claudi}, {Lucatello},
  \& {Barbieri}}]{gratton03}
{Gratton}, R.~G., {Carretta}, E., {Claudi}, R., {Lucatello}, S., \& {Barbieri},
  M. 2003, \aap, 404, 187

\bibitem[{{H{\" a}nninen} \& {Flynn}(2002)}]{hanninen02}
{H{\" a}nninen}, J., \& {Flynn}, C. 2002, \mnras, 337, 731

\bibitem[{{Haywood}(2001)}]{haywood01}
{Haywood}, M. 2001, \mnras, 325, 1365

\bibitem[{{Holland} {et~al.}(1996){Holland}, {Fahlman}, \&
  {Richer}}]{holland96}
{Holland}, S., {Fahlman}, G.~G., \& {Richer}, H.~B. 1996, \aj, 112, 1035

\bibitem[{{Hummel} \& {Dettmar}(1990)}]{hummel90}
{Hummel}, E., \& {Dettmar}, R.-J. 1990, \aap, 236, 33

\bibitem[{{Hummel} {et~al.}(1986){Hummel}, {Dettmar}, \&
  {Wielebinski}}]{hummel86}
{Hummel}, E., {Dettmar}, R.-J., \& {Wielebinski}, R. 1986, \aap, 166, 97

\bibitem[{{Jenkins}(1992)}]{jenkins92}
{Jenkins}, A. 1992, \mnras, 257, 620

\bibitem[{{Jenkins} \& {Binney}(1990)}]{jenkins90}
{Jenkins}, A., \& {Binney}, J. 1990, \mnras, 245, 305

\bibitem[{{Kravtsov} {et~al.}(1997){Kravtsov}, {Ipatov}, {Samus}, {Smirnov},
  {Alcaino}, {Liller}, \& {Alvarado}}]{kravtsov97}
{Kravtsov}, V., {Ipatov}, A., {Samus}, N., {Smirnov}, O., {Alcaino}, G.,
  {Liller}, W., \& {Alvarado}, F. 1997, \aaps, 125, 1

\bibitem[{{Kroupa}(2002)}]{kroupa02}
{Kroupa}, P. 2002, \mnras, 330, 707

\bibitem[{{Kroupa} {et~al.}(1993){Kroupa}, {Tout}, \& {Gilmore}}]{kroupa93}
{Kroupa}, P., {Tout}, C.~A., \& {Gilmore}, G. 1993, \mnras, 262, 545

\bibitem[{{L{\' o}pez-Corredoira} {et~al.}(2002){L{\' o}pez-Corredoira},
  {Cabrera-Lavers}, {Garz{\' o}n}, \& {Hammersley}}]{lopez02}
{L{\' o}pez-Corredoira}, M., {Cabrera-Lavers}, A., {Garz{\' o}n}, F., \&
  {Hammersley}, P.~L. 2002, \aap, 394, 883

\bibitem[{{Lacey}(1991)}]{lacey91}
{Lacey}, C.~G. 1991, in Dynamics of Disc Galaxies, Proceedings of Varberg
  Conference, ed. B. Sundelius, 257--+

\bibitem[{{Lacey} \& {Ostriker}(1985)}]{lacey85}
{Lacey}, C.~G., \& {Ostriker}, J.~P. 1985, \apj, 299, 633

\bibitem[{{Lee} {et~al.}(1993){Lee}, {Freedman}, \& {Madore}}]{lee93}
{Lee}, M.~G., {Freedman}, W.~L., \& {Madore}, B.~F. 1993, \apj, 417, 553

\bibitem[{Leroy {et~al.}(2005)Leroy, Bolatto, Simon, \& Blitz}]{leroy05}
Leroy, A., Bolatto, A.~D., Simon, J.~D., \& Blitz, L. 2005

\bibitem[{{Marigo}(2001)}]{marigo01}
{Marigo}, P. 2001, \aap, 370, 194

\bibitem[{{Martin}(1998)}]{martin98}
{Martin}, M.~C. 1998, \aaps, 131, 77

\bibitem[{{Matthews} \& {Wood}(2001)}]{matthews01}
{Matthews}, L.~D., \& {Wood}, K. 2001, \apj, 548, 150

\bibitem[{{Mendez} \& {van Altena}(1998)}]{mendez98}
{Mendez}, R.~A., \& {van Altena}, W.~F. 1998, \aap, 330, 910

\bibitem[{{Mishenina} {et~al.}(2004){Mishenina}, {Soubiran}, {Kovtyukh}, \&
  {Korotin}}]{mishenina04}
{Mishenina}, T.~V., {Soubiran}, C., {Kovtyukh}, V.~V., \& {Korotin}, S.~A.
  2004, \aap, 418, 551

\bibitem[{{Mould}(2005)}]{mould05}
{Mould}, J. 2005, \aj, 129, 698

\bibitem[{{Narayan} \& {Jog}(2002)}]{narayan02}
{Narayan}, C.~A., \& {Jog}, C.~J. 2002, \aap, 390, L35

\bibitem[{{Neeser} {et~al.}(2002){Neeser}, {Sackett}, {De Marchi}, \&
  {Paresce}}]{neeser02}
{Neeser}, M.~J., {Sackett}, P.~D., {De Marchi}, G., \& {Paresce}, F. 2002,
  \aap, 383, 472

\bibitem[{{Ng} {et~al.}(1996){Ng}, {Bertelli}, {Chiosi}, \& {Bressan}}]{ng96}
{Ng}, Y.~K., {Bertelli}, G., {Chiosi}, C., \& {Bressan}, A. 1996, \aap, 310,
  771

\bibitem[{{Nordstr{\" o}m} {et~al.}(2004){Nordstr{\" o}m}, {Mayor}, {Andersen},
  {Holmberg}, {Pont}, {J{\o}rgensen}, {Olsen}, {Udry}, \&
  {Mowlavi}}]{nordstrom04}
{Nordstr{\" o}m}, B., {Mayor}, M., {Andersen}, J., {Holmberg}, J., {Pont}, F.,
  {J{\o}rgensen}, B.~R., {Olsen}, E.~H., {Udry}, S., \& {Mowlavi}, N. 2004,
  \aap, 418, 989

\bibitem[{{Ojha}(2001)}]{ojha01}
{Ojha}, D.~K. 2001, \mnras, 322, 426

\bibitem[{{Olling}(1996)}]{olling96}
{Olling}, R.~P. 1996, \aj, 112, 457

\bibitem[{{Pagel}(1997)}]{pagel97}
{Pagel}, B.~E.~J. 1997, {Nucleosynthesis and chemical evolution of galaxies}
  (Nucleosynthesis and chemical evolution of galaxies
  /B.~E.~J.~Pagel.~Cambridge : Cambridge University Press, 1997.~ ISBN
  0521550610)

\bibitem[{{Paturel} {et~al.}(2003){Paturel}, {Theureau}, {Bottinelli},
  {Gouguenheim}, {Coudreau-Durand}, {Hallet}, \& {Petit}}]{paturel03}
{Paturel}, G., {Theureau}, G., {Bottinelli}, L., {Gouguenheim}, L.,
  {Coudreau-Durand}, N., {Hallet}, N., \& {Petit}, C. 2003, \aap, 412, 57

\bibitem[{{Paturel} {et~al.}(1995){Paturel}, {Vauglin}, {Andernach}, {Garnier},
  {Marthinet}, {Petit}, {di Nella}, {Bottinelli}, {Gouguenheim}, \&
  {Durand}}]{paturel95}
{Paturel}, G., {Vauglin}, I., {Andernach}, H., {Garnier}, R., {Marthinet},
  M.-C., {Petit}, C., {di Nella}, H., {Bottinelli}, L., {Gouguenheim}, L., \&
  {Durand}, N. 1995, {LEDA: The Lyon-Meudon Extragalactic Database} (ASSL
  Vol.~203: Information On-Line Data in Astronomy), 115--126

\bibitem[{{Pohlen} {et~al.}(2004){Pohlen}, {Balcells}, {L{\" u}tticke}, \&
  {Dettmar}}]{pohlen04}
{Pohlen}, M., {Balcells}, M., {L{\" u}tticke}, R., \& {Dettmar}, R.-J. 2004,
  \aap, 422, 465

\bibitem[{{Pohlen} {et~al.}(2000){Pohlen}, {Dettmar}, {L{\" u}tticke}, \&
  {Schwarzkopf}}]{pohlen00}
{Pohlen}, M., {Dettmar}, R.-J., {L{\" u}tticke}, R., \& {Schwarzkopf}, U. 2000,
  \aaps, 144, 405

\bibitem[{{Quinn} {et~al.}(1993){Quinn}, {Hernquist}, \& {Fullagar}}]{quinn93}
{Quinn}, P.~J., {Hernquist}, L., \& {Fullagar}, D.~P. 1993, \apj, 403, 74

\bibitem[{{Sarajedini} \& {Van Duyne}(2001)}]{sarajedini01}
{Sarajedini}, A., \& {Van Duyne}, J. 2001, \aj, 122, 2444

\bibitem[{{Schmidt}(1963)}]{schmidt63}
{Schmidt}, M. 1963, \apj, 137, 758

\bibitem[{{Sellwood} \& {Carlberg}(1984)}]{sellwood84}
{Sellwood}, J.~A., \& {Carlberg}, R.~G. 1984, \apj, 282, 61

\bibitem[{{Seth} {et~al.}(2005){Seth}, {Dalcanton}, \& {de Jong}}]{seth05}
{Seth}, A.~C., {Dalcanton}, J.~J., \& {de Jong}, R.~S. 2005, \aj, 129, 1331

\bibitem[{{Shapiro} {et~al.}(2003){Shapiro}, {Gerssen}, \& {van der
  Marel}}]{shapiro03}
{Shapiro}, K.~L., {Gerssen}, J., \& {van der Marel}, R.~P. 2003, \aj, 126, 2707

\bibitem[{{Siegel} {et~al.}(2002){Siegel}, {Majewski}, {Reid}, \&
  {Thompson}}]{siegel02}
{Siegel}, M.~H., {Majewski}, S.~R., {Reid}, I.~N., \& {Thompson}, I.~B. 2002,
  \apj, 578, 151

\bibitem[{{Spitzer} \& {Schwarzschild}(1951)}]{spitzer51}
{Spitzer}, L.~J., \& {Schwarzschild}, M. 1951, \apj, 114, 385

\bibitem[{{Tiede} {et~al.}(2004){Tiede}, {Sarajedini}, \& {Barker}}]{tiede04}
{Tiede}, G.~P., {Sarajedini}, A., \& {Barker}, M.~K. 2004, \aj, 128, 224

\bibitem[{Tikhonov \& Galazutdinova(2005)}]{tikhonov05b}
Tikhonov, N.~A., \& Galazutdinova, O.~A. 2005

\bibitem[{{Tikhonov} {et~al.}(2005){Tikhonov}, {Galazutdinova}, \&
  {Drozdovsky}}]{tikhonov05a}
{Tikhonov}, N.~A., {Galazutdinova}, O.~A., \& {Drozdovsky}, I.~O. 2005, \aap,
  431, 127

\bibitem[{{Truran} \& {Cameron}(1971)}]{truran71}
{Truran}, J.~W., \& {Cameron}, A.~G.~W. 1971, \apss, 14, 179

\bibitem[{{Tsikoudi}(1979)}]{tsikoudi79}
{Tsikoudi}, V. 1979, \apj, 234, 842

\bibitem[{{van der Kruit}(1988)}]{vanderkruit88}
{van der Kruit}, P.~C. 1988, \aap, 192, 117

\bibitem[{{van der Kruit} \& {Searle}(1981)}]{vanderkruit81}
{van der Kruit}, P.~C., \& {Searle}, L. 1981, \aap, 95, 105

\bibitem[{{Verde} {et~al.}(2002){Verde}, {Oh}, \& {Jimenez}}]{verde02}
{Verde}, L., {Oh}, S.~P., \& {Jimenez}, R. 2002, \mnras, 336, 541

\bibitem[{{Wielen}(1977)}]{wielen77}
{Wielen}, R. 1977, \aap, 60, 263

\bibitem[{{Wyse} \& {Gilmore}(1995)}]{wyse95}
{Wyse}, R.~F.~G., \& {Gilmore}, G. 1995, \aj, 110, 2771

\bibitem[{Yoachim \& Dalcanton(2005)}]{yoachim05}
Yoachim, P., \& Dalcanton, J.~J. 2005

\bibitem[{{Young} \& {Scoville}(1991)}]{young91}
{Young}, J.~S., \& {Scoville}, N.~Z. 1991, \araa, 29, 581

\bibitem[{{Zibetti} {et~al.}(2004){Zibetti}, {White}, \&
  {Brinkmann}}]{zibetti04a}
{Zibetti}, S., {White}, S.~D.~M., \& {Brinkmann}, J. 2004, \mnras, 347, 556

\end{thebibliography}
\end{document}